\documentclass[a4paper,fleqn,usenatbib]{mnras}

\usepackage{newtxtext,newtxmath}

\usepackage[T1]{fontenc}
\usepackage{ae,aecompl}

\usepackage{graphicx}	
\usepackage{amsmath}	
\usepackage{amssymb}	
\usepackage{epstopdf}
\usepackage{multirow}
\usepackage{ulem}
\usepackage{xspace}

\newcommand\arcse{\hbox{$^{\prime\prime}$}\xspace}   
\newcommand*{\SAs}{{Sgr\,A$^*$}\xspace}

\title[Modelling X-rays around the Galactic Centre]{Modelling the thermal X-ray emission around the Galactic Centre from colliding Wolf-Rayet winds}

\author[C.~M.~P.~Russell, Q.~D.~Wang, and J.~Cuadra]{
Christopher M. P. Russell,$^{1}$\thanks{E-mail: crussell@udel.edu (CMPR)}
Q. Daniel Wang,$^{2}$
and Jorge Cuadra$^{3}$
\\
$^{1}$X-ray Astrophysics Laboratory, Code 662, NASA/Goddard Space Flight Center, Greenbelt, MD 20771, USA\\
$^{2}$Department of Astronomy, University of Massachusetts, Amherst, MA 01003, USA\\
$^{3}$Instituto de Astrof\'{\i}sica, Facultad de F\'{\i}sica, Pontificia Universidad Cat\'{o}lica de Chile, 782-0436 Santiago, Chile
}

\date{Accepted 2016 October 5. Received 2016 October 4; in original form 2016 January 29}

\pubyear{2017}

\begin{document}
\label{firstpage}
\pagerange{\pageref{firstpage}--\pageref{lastpage}}
\maketitle

\begin{abstract}
  The Galactic Centre is a hotbed of astrophysical activity, with the injection of wind material from $\sim$30 massive Wolf--Rayet (WR) stars orbiting within 12 arcsec of the supermassive black hole (SMBH) playing an important role.
  Hydrodynamic simulations of such colliding and accreting winds produce a complex density and temperature structure of cold wind material shocking with the ambient medium, creating a large reservoir of hot, X-ray-emitting gas.
  This work aims to confront the 3 Ms of \textit{Chandra} X-ray Visionary Program observations of this diffuse emission by computing the X-ray emission from these hydrodynamic simulations of the colliding WR winds, amid exploring a variety of SMBH feedback mechanisms.
  The major success of the model is that it reproduces the spectral shape from the 2--5 arcsec ring around the SMBH, where most of the stellar wind material that is ultimately captured by Sgr A* is shock-heated and thermalized.
  This naturally explains that the hot gas comes from colliding WR winds, and that the wind speeds of these stars are, in general, well constrained.
  The flux level of these spectra, as well as 12 $\times$ 12-arcsec$^2$ images of 4--9 keV, shows that the X-ray flux is tied to the SMBH feedback strength; stronger feedback clears out more hot gas, thereby decreasing the thermal X-ray emission.
  The model in which \SAs produced an intermediate-strength outflow during the last few centuries best matches the observations to within about 10 per cent, showing that SMBH feedback is required to interpret the X-ray emission in this region.
\end{abstract}




\section{Introduction}

The Galactic Centre is a hotbed of astrophysical activity, hosting myriad stars of varying masses orbiting a supermassive black hole (SMBH) of $\sim$3.5$\times10^6$\,M$_\odot$ \citep[e.g.][]{GhezP05} associated with the radio source \SAs.  Classified as an inactive SMBH due to its low accretion rate \citep*{MeliaFalcke01,GenzelEisenhauerGillessen10}, the proximity of \SAs makes it the only SMBH where its orbiting stars are resolved, and thus the only instance where the stellar, wind, and orbital parameters of the individual stars orbiting an SMBH can be accurately determined.  Therefore, the Galactic Centre provides the best opportunity to study the interplay between an SMBH and the stars and ejected wind material orbiting it.

To study the accretion rate on to \SAs, \citet{CuadraNayakshinMartins08} computed hydrodynamic simulations of the winds of 30 Wolf--Rayet (WR) stars (evolved massive stars with the highest mass-loss rates around the Galactic Centre) that orbit within $\sim$10 arcsec (=\,0.4\,pc at a distance of 8.25kpc) of \SAs.  The simulations ran from 1100 yr ago to the present day, and are based on orbital data from \citet{PaumardP06} and \citet{BeloborodovP06} and stellar wind data from \citet{MartinsP07}.  The end result predicted the time-dependent accretion rate of material on to \SAs while also producing the density and temperature structure of the hot/shocked and the cold/unshocked wind material ejected from the WR stars.  Although the simulation does not include any mass ejected from stars with lower mass-loss rates, like the O stars, nor any material ejected prior to 1100 yr ago, it is the most complete calculation of the material around the Galactic Centre out to $\sim$10 arcsec.

X-ray observations of the Galactic Centre provide key insights into the region as these high-energy photons can penetrate the high absorption column through the Galactic plane.  The \textit{Chandra} X-ray Visionary Program (XVP) on the Galactic Centre, which performed grating observations of the region for 3 Ms during both flaring and quiescent states, showed that the X-ray properties of \SAs indicate that it is a radiatively inefficient accretion flow (RIAF) and has an outflow \citep{WangP13}.  \textit{XMM--Newton} observations of several 100 pc away from the Galactic Centre revealed that the X-ray activity, and by extension the total activity, of \SAs was higher several centuries ago \citep{PontiP10}.  From these two results, \citet*{CuadraNayakshinWang15} ran more simulations to account for possible SMBH feedback mechanisms over a range of feedback strengths.
As expected, these different feedback mechanisms altered the dynamics of the colliding WR winds, significantly modifying the density and temperature structure around \SAs in the models with the strongest feedback.

In addition to studying the point-source emission of the SMBH, the XVP also provided the best observations of the spatially and spectrally resolved X-ray emission out to $\sim$20 arcsec from \SAs.  This diffuse emission is thermal, so it is thought to originate from the colliding stellar winds of the stars orbiting the SMBH.  Therefore, this observation set provides an excellent test for the validity of the hydrodynamic simulations, as well as a means to distinguish between feedback models.

This work computes the thermal X-ray emission from the aforementioned hydrodynamics simulations and compares the results to the observations with the aim of increasing our understanding of the WR stars, and more generally the full environment, surrounding \SAs.
Section~\ref{M} recaps the relevant details of the hydrodynamic simulations, describes the method for computing the X-ray emission from these simulations, and discusses the observed X-ray image and spectrum that the modelling aims to explain.
Section~\ref{R} presents the results of the X-ray calculations and compares them with the observations. We discuss the results in Section~\ref{D} and present our conclusions in Section~\ref{C}.

\section{Method}\label{M}

\subsection{Hydrodynamics}\label{H}

The hydrodynamic simulations presented in this work are all from \citet{CuadraNayakshinMartins08,CuadraNayakshinWang15}.  We briefly mention the features important to this work here; the reader is referred to these works for the full details of the simulations.

These computations used the smoothed particle hydrodynamics (SPH) code \textsc{gadget-2} \citep{Springel05} to follow the orbits of the 30 WR stars around \SAs, all the while ejecting their wind material into the simulation volume.  The SMBH particle is a sink particle \citep*{BateBonnellPrice95}, which accretes any gas particle that comes within 0.1\arcse of \SAs (due to non-relativistic and resolution requirements) by removing it from the simulation, as implemented by \citet{CuadraP06}.  The WR stars are source particles, which inject the standard SPH particles to represent their winds.  All the winds shock very far from their stellar surfaces, so the gas particles are injected at their terminal speed instead of being accelerated away from their star.  To bind the IRS~13E cluster, which contains two WR stars, a 350-M$_\odot$ dark matter particle is included in the cluster.  The winds range in mass-loss rate from $\dot{M}=5\times10^{-5}$ to $5\times10^{-4}$\,M$_\odot$\,yr$^{-1}$ and in velocity from $v=600$ to $2500$\,km\,s$^{-1}$.
For the stellar orbits, we are using the 1disc model of \citet{CuadraNayakshinMartins08}, which has approximately half the stars in a disc \citep{BeloborodovP06} and the remaining stars in a more isotropic distribution, which is consistent with the observations of individual stellar orbits by \citet{LuP09}.

\citet{CuadraNayakshinMartins08} performed the SPH simulation with no feedback (NF) from the SMBH, which acts as a control model. \citet{CuadraNayakshinWang15} performed many more simulations with various feedbacks.  The outflow model (OF) takes every SPH particle that enters the 0.1-arcsec sphere near \SAs and ejects it back into the simulation domain (so $\dot{M}_\textrm{out}=\dot{M}_\textrm{accrete}$) in a random direction with $v=10\,000$\,km/s.  This is because less than 1 per cent of the material entering the Bondi radius of the RIAF SMBH actually accretes \citep[e.g.][]{BlandfordBegelman99,WangP13}, so it is convenient to simply expel every particle that enters the SMBH domain.  Three outburst models attempt to account for the inferred larger activity of \SAs in the past by ejecting a significant amount of material and energy into the simulation domain at the SMBH boundary from 400 to 100 yr ago.  Before and after this time interval, the simulation is just like the NF model.  Two spherical outburst models have $\dot{M}_\textrm{out}=10^{-4}$M$_\odot$\,yr$^{-1}$ and outflow velocities of $v=5000$ and $10\,000$\,km s$^{-1}$ (OB5 and OB10, respectively).  The third outburst model has a bipolar flow (OBBP) confined to a cone with a half-opening angle of 15$^\circ$.  Its outflow velocity is $v=5000$\,km s$^{-1}$.

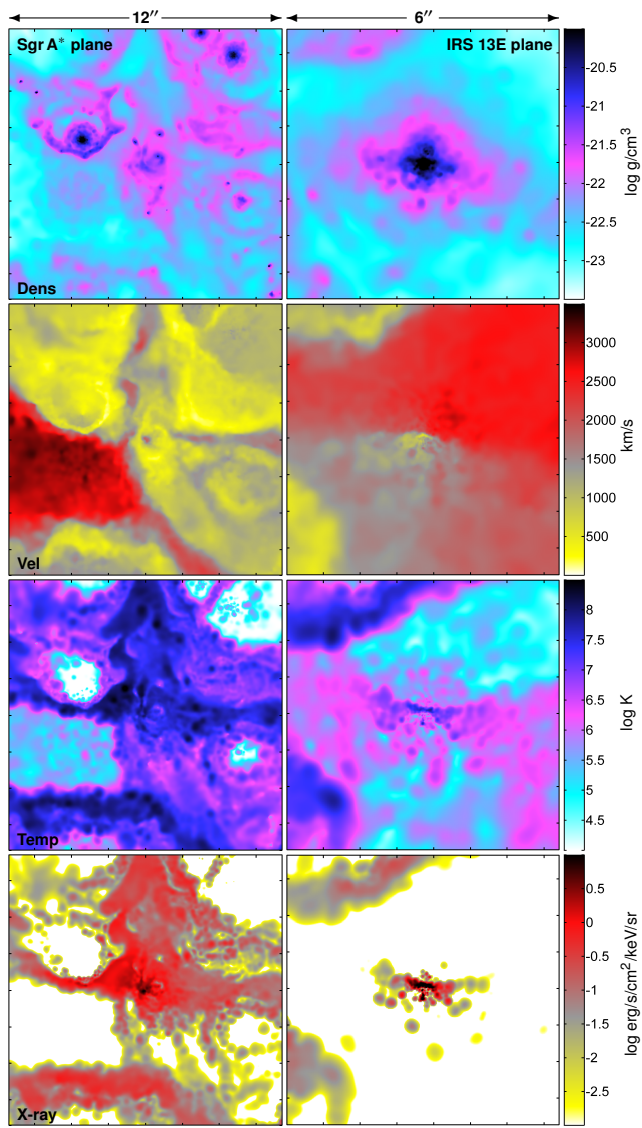
\begin{figure}
  \includegraphics[height=3.65cm]{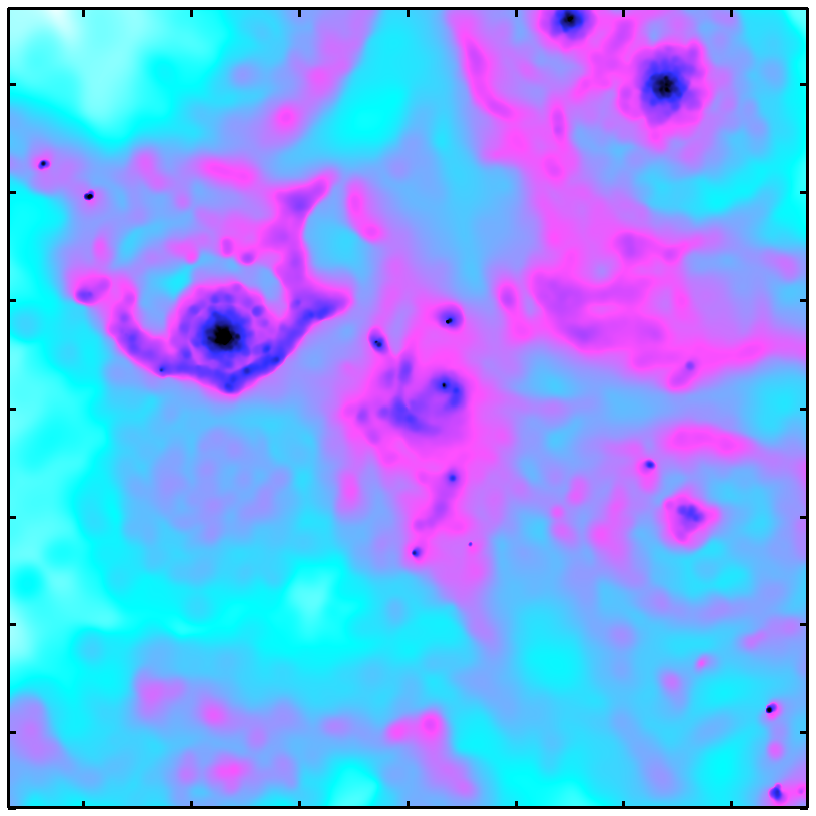}
  \put(-100,95){\scriptsize \fontfamily{phv}\selectfont \textbf{\SAs plane}}
  \put(-100,3){\scriptsize \fontfamily{phv}\selectfont \textbf{Dens}}
  \put(-58,107){\vector(-1,0){45}}
  \put(-46,107){\vector(1,0){45}}
  \put(-57,105){\scriptsize \fontfamily{phv}\selectfont \textbf{12\arcsec}}
  \includegraphics[height=3.65cm]{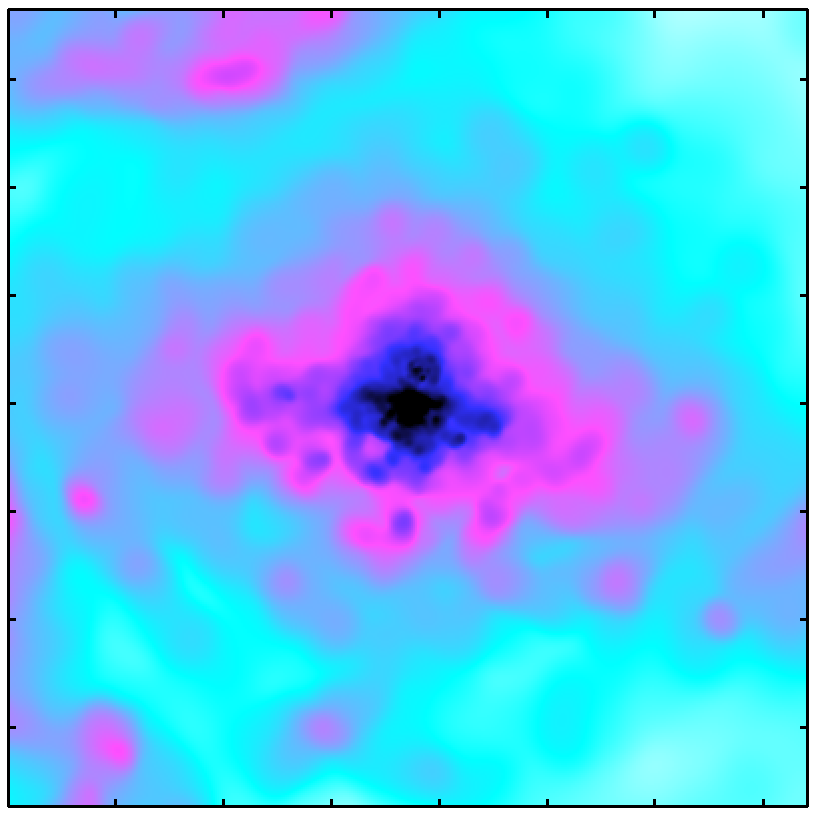}
  \put(-43,95){\scriptsize \fontfamily{phv}\selectfont \textbf{IRS~13E plane}}
  \put(-58,107){\vector(-1,0){45}}
  \put(-46,107){\vector(1,0){45}}
  \put(-55,105){\scriptsize \fontfamily{phv}\selectfont \textbf{6\arcsec}}
  \includegraphics[height=3.65cm]{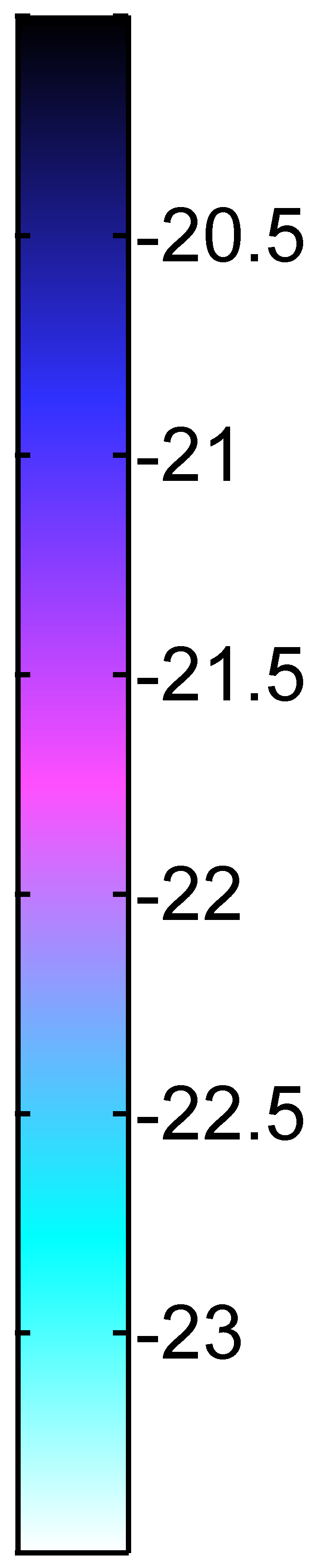}\put(1,40){\scriptsize \fontfamily{phv}\selectfont \rotatebox{90}{log g/cm\textsuperscript{\tiny 3}}}

  \vspace{-0.033cm}

  \includegraphics[height=3.65cm]{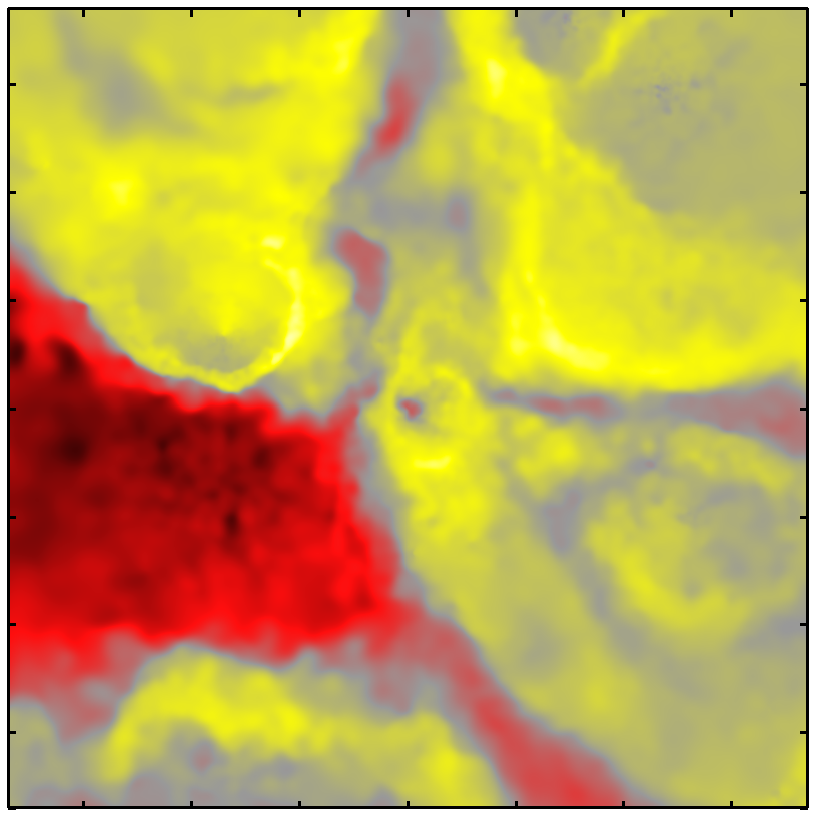}\put(-100,3){\scriptsize \fontfamily{phv}\selectfont \textbf{Vel}}%
  \includegraphics[height=3.65cm]{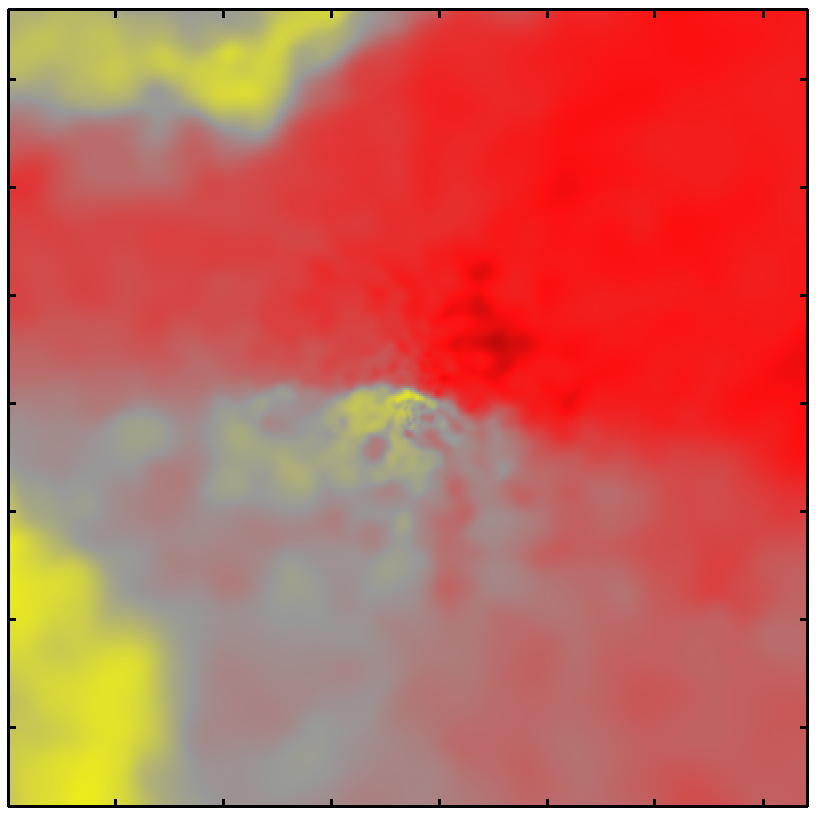}%
  \includegraphics[height=3.65cm]{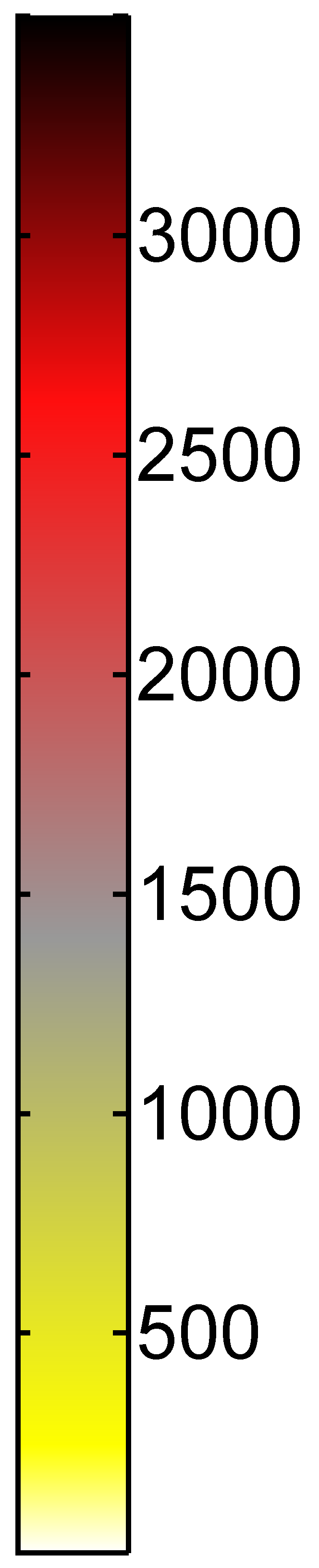}\put(1,47){\scriptsize \fontfamily{phv}\selectfont \rotatebox{90}{km/s}}

  \vspace{-0.033cm}

  \includegraphics[height=3.65cm]{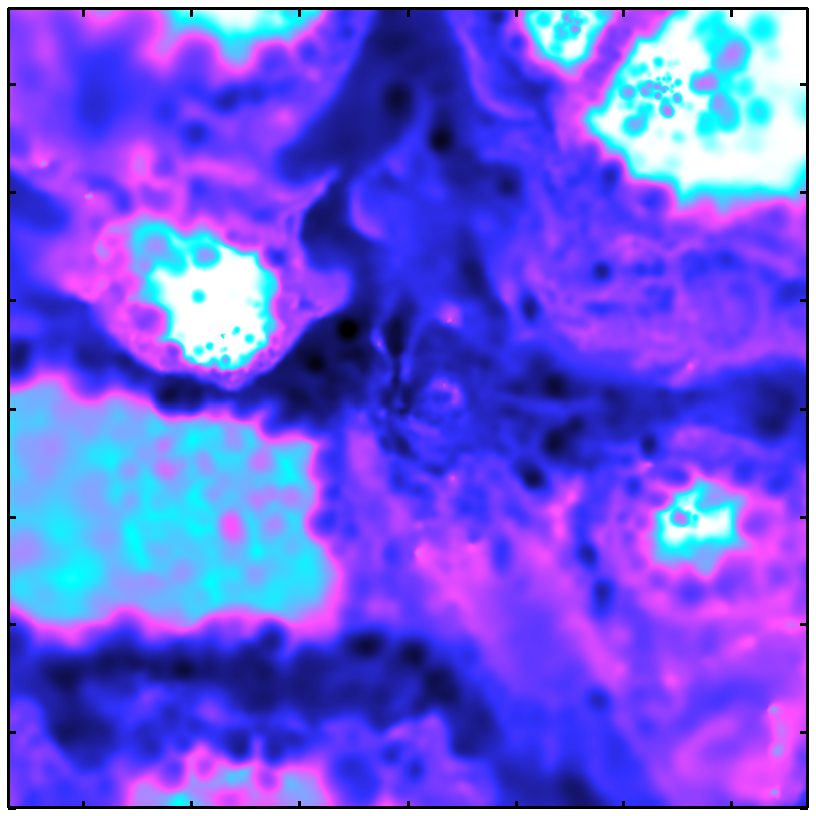}\put(-100,3){\scriptsize \fontfamily{phv}\selectfont \textbf{Temp}}%
  \includegraphics[height=3.65cm]{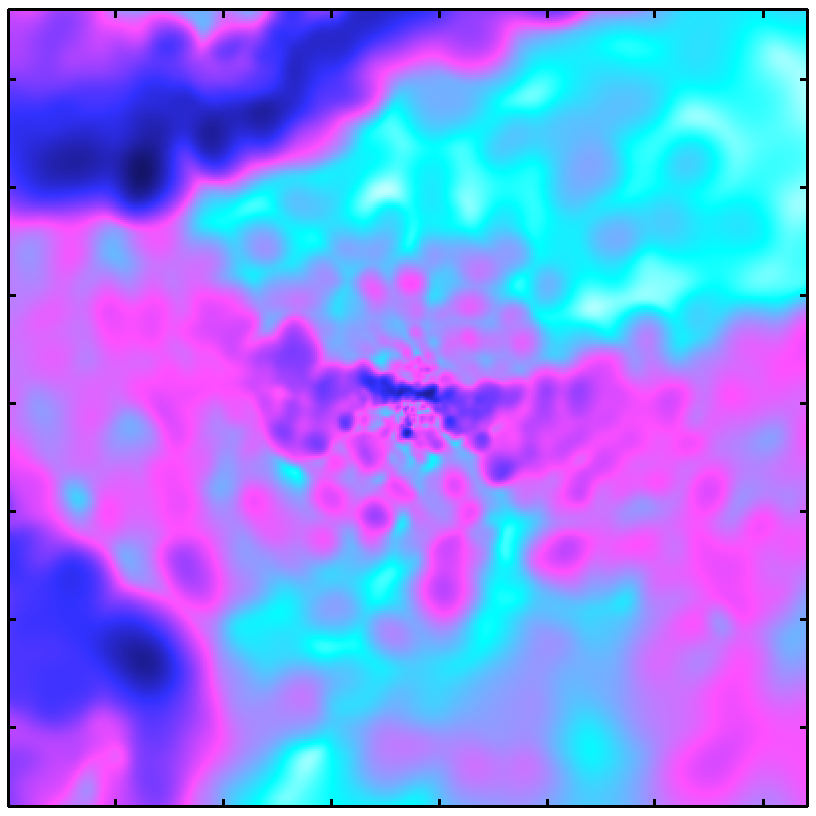}%
  \includegraphics[height=3.65cm]{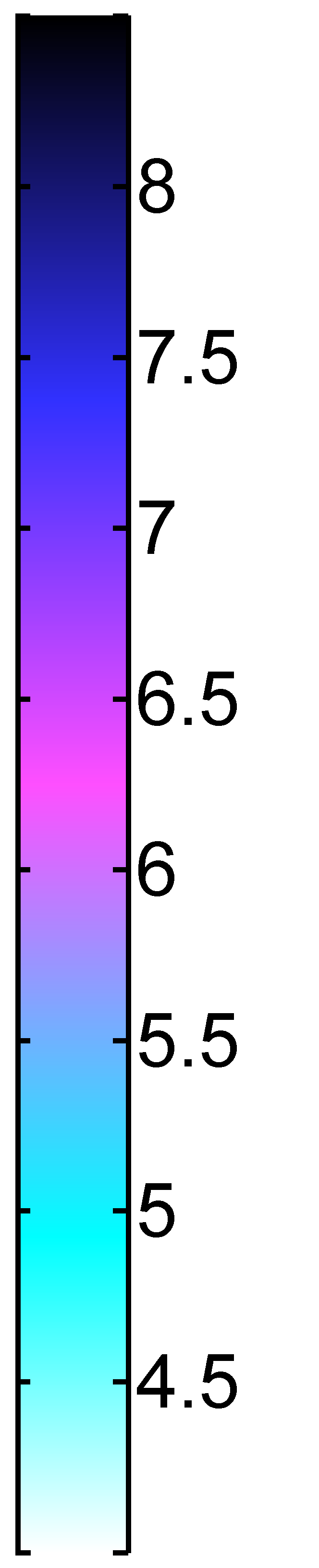}\put(1,46){\scriptsize \fontfamily{phv}\selectfont \rotatebox{90}{log K}}

  \vspace{-0.033cm}

  \includegraphics[height=3.65cm]{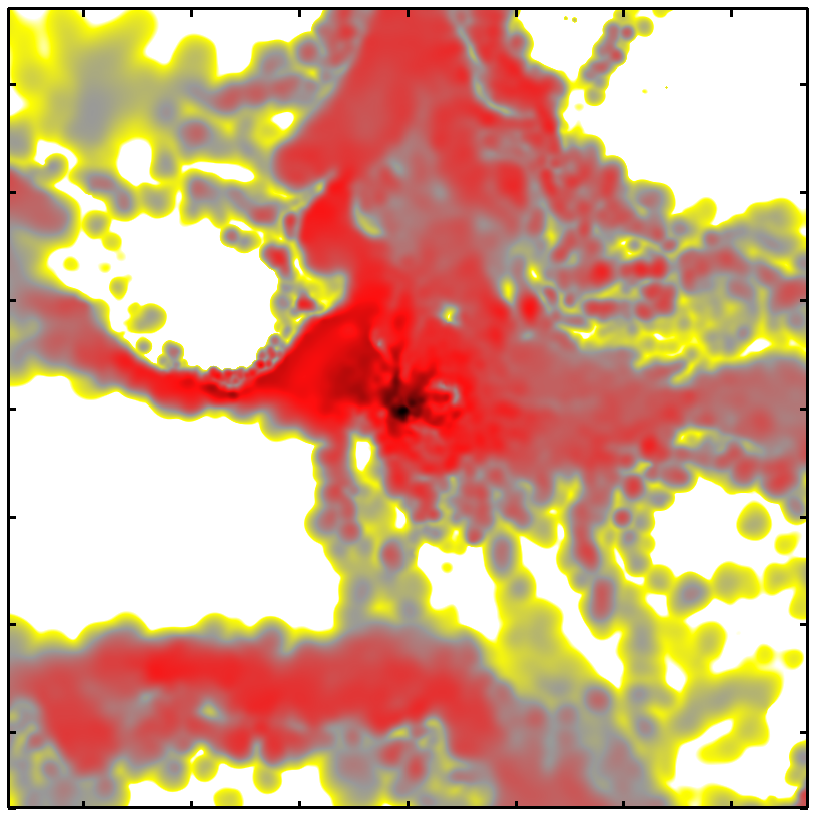}\put(-100,3){\scriptsize \fontfamily{phv}\selectfont \textbf{X-ray}}%
  \includegraphics[height=3.65cm]{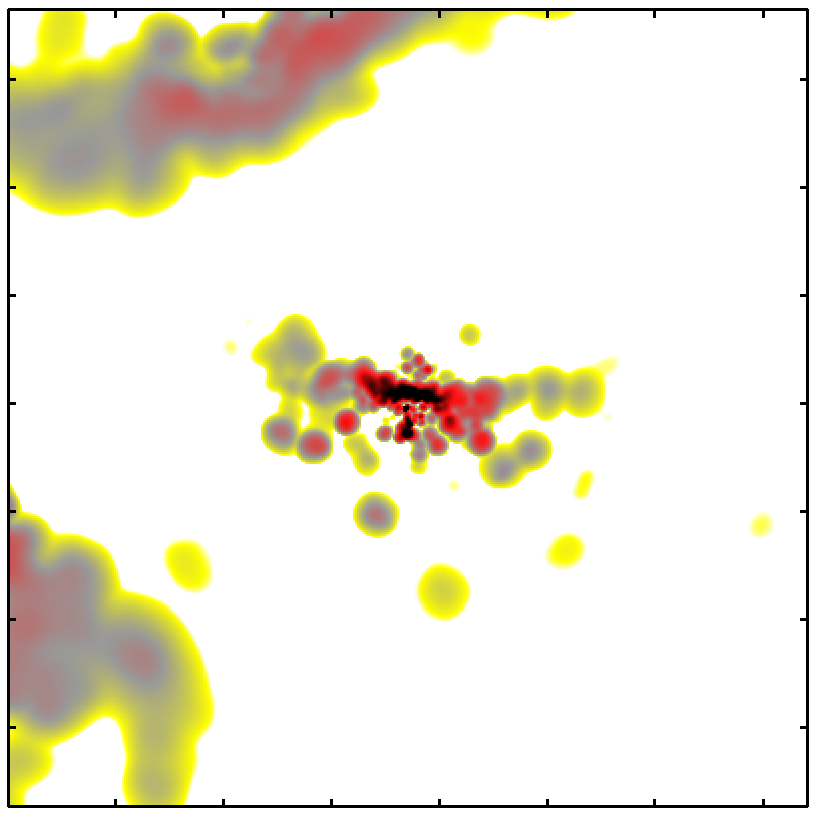}%
  \includegraphics[height=3.65cm]{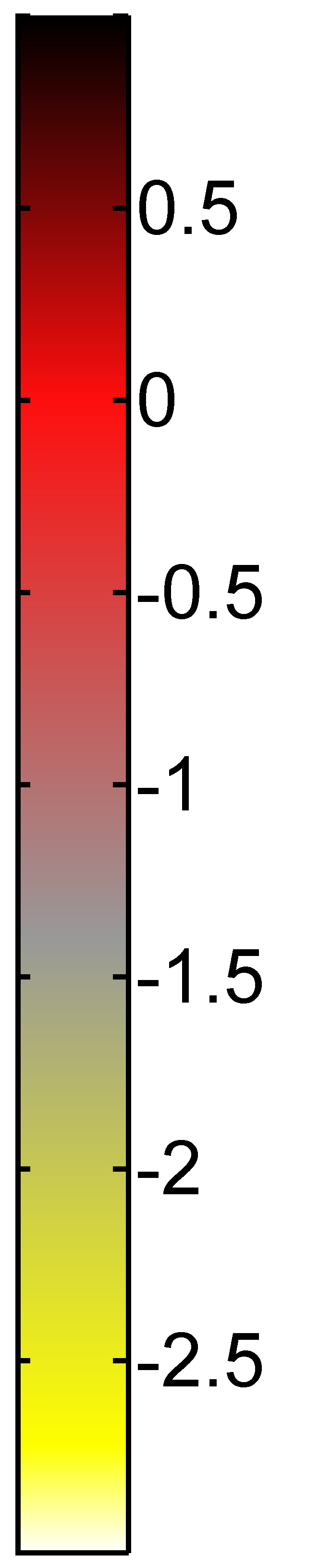}\put(1,27){\scriptsize \fontfamily{phv}\selectfont \rotatebox{90}{log erg/s/cm\textsuperscript{\tiny 2}/keV/sr}}

  \caption{Panels of NF showing (from top to bottom) the density, speed, temperature, and 6-keV X-ray source function.  The left-hand column shows 12 $\times$ 12 arcsec$^2$ at the midplane centred on the SMBH and has tick marks at every 2$\times$10$^{17}$ cm, while the right-hand column shows the 6 $\times$ 6 arcsec$^2$ region centred on IRS~13E and has tick marks at every 10$^{17}$ cm.  In the mid-plane (left-hand column), three WRs happen to be located nearby (one in the upper-middle left-hand portion and the other two in the upper right-hand portion of the panels; most easily identified in the density panel), and so shows the high-speed, low-temperature regions of unshocked material colliding into the ambient medium to create hot, X-ray-producing gas. The IRS~13E plane (right-hand column) shows the strong X-ray emission from between this cluster's two WRs.}
  \label{fi:SPH}
\end{figure}

Fig.~\ref{fi:SPH} shows the density, speed, temperature, and X-ray source function at 6 keV (described in Section~\ref{RT}) of the NF model at two planes perpendicular to our line of sight.  The mid-plane images (left-hand column) show three WRs interacting with the surrounding medium.  The region in the immediate vicinity is filled with cool material flowing at its respective wind terminal speed, until it shocks with previously ejected material and heats up to X-ray-generating temperatures.  The projected motion of the star on the left-hand side is predominantly down at this instant, so it forms a bow shock (which projects to an arc in 2D) along its direction of motion opening up around the WR.  The low-temperature cavities and the hot, X-ray-generating gas are also seen around the other two WRs in the upper right-hand side of the panels.  The IRS~13E cluster (right-hand column) clearly shows a strong wind--wind collision region between the two WRs, which lights up very brightly in X-rays, more so than the rest of the simulation volume.  This is due to the closeness of these shocks to their WRs, thus creating a situation of higher shock density than elsewhere around \SAs.

\begin{figure*}
  \begin{flushleft}

  \includegraphics[height=3.34cm]{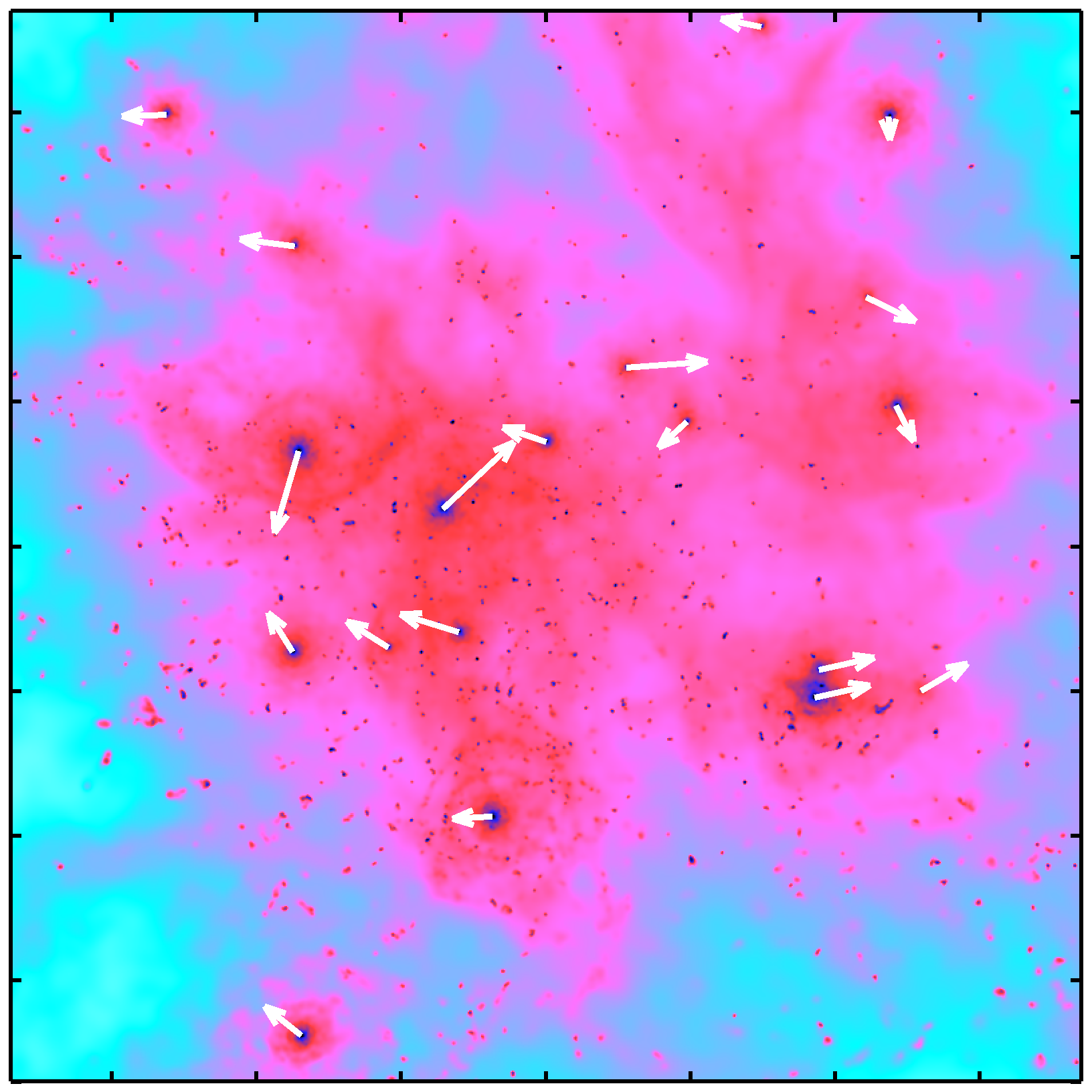}
  \put(-93,3){\scriptsize \fontfamily{phv}\selectfont \textbf{NF}}
  \put(-35,28){\scriptsize \fontfamily{phv}\selectfont \textbf{IRS~13E}}
  \put(-54,98){\vector(-1,0){40}}
  \put(-41,98){\vector(1,0){40}}
  \put(-52,96){\scriptsize \fontfamily{phv}\selectfont \textbf{12\arcsec}}
  \includegraphics[height=3.34cm]{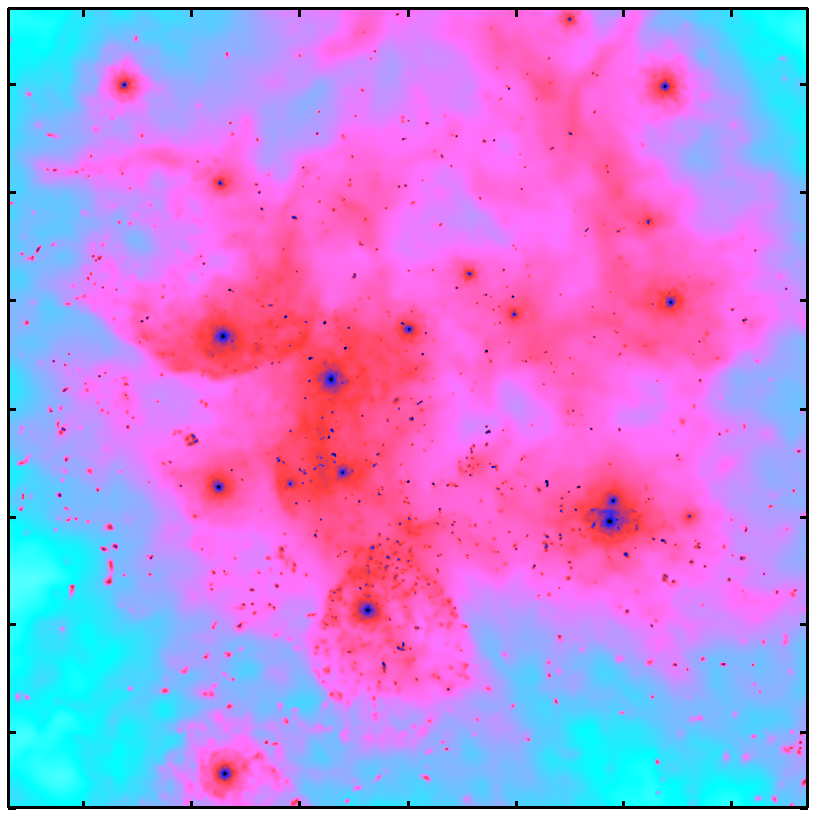}\put(-93,3){\scriptsize \fontfamily{phv}\selectfont \textbf{OF}}%
  \includegraphics[height=3.34cm]{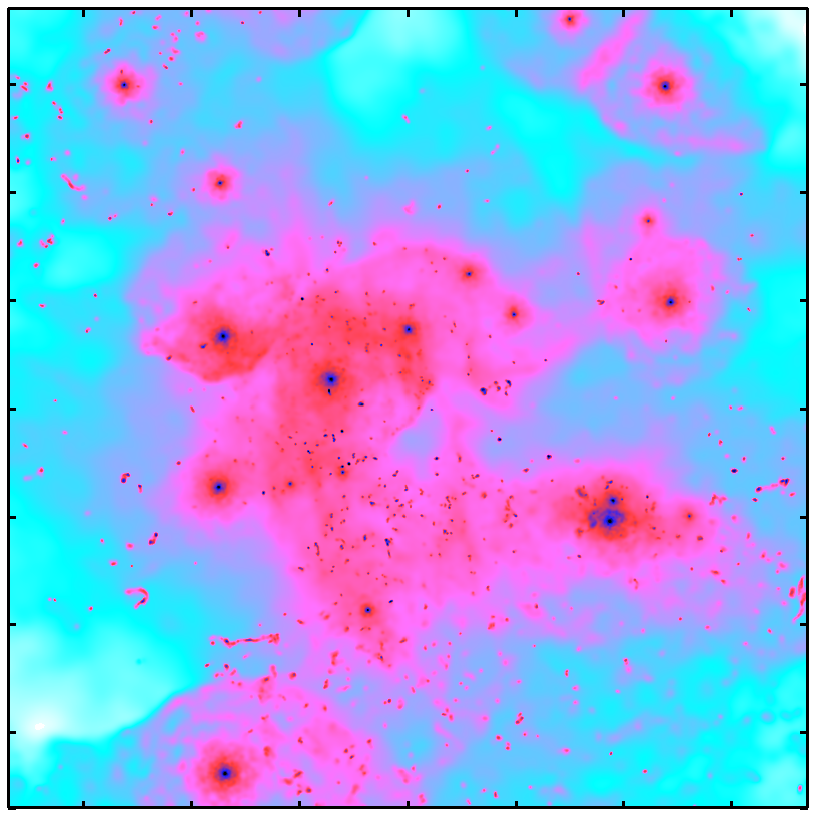}\put(-93,3){\scriptsize \fontfamily{phv}\selectfont \textbf{OBBP}}%
  \includegraphics[height=3.34cm]{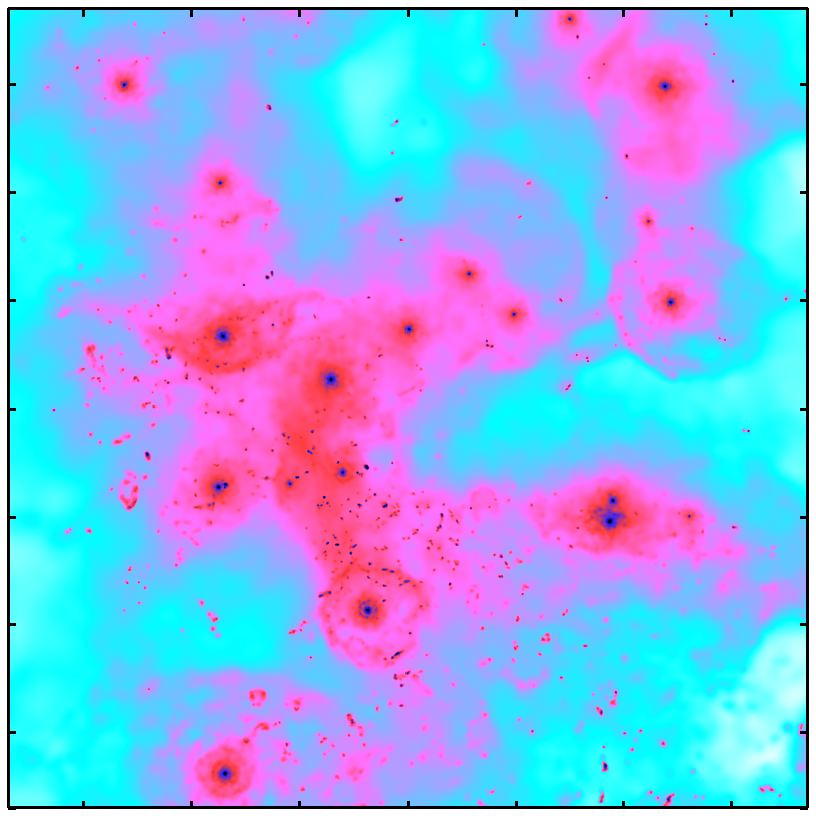}\put(-93,3){\scriptsize \fontfamily{phv}\selectfont \textbf{OB5}}%
  \includegraphics[height=3.34cm]{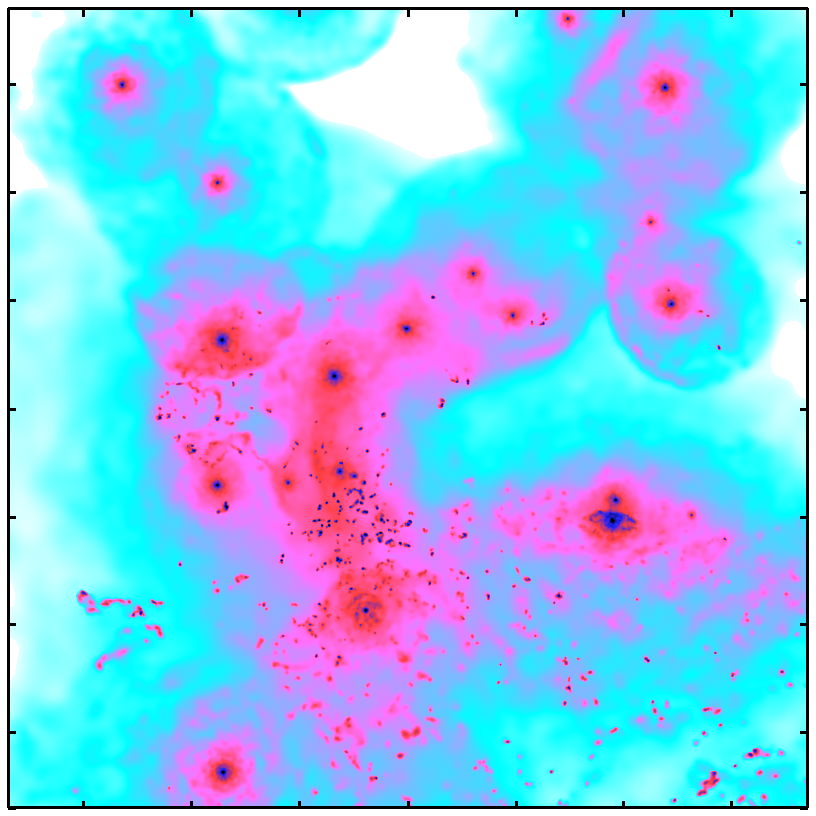}\put(-93,3){\scriptsize \fontfamily{phv}\selectfont \textbf{OB10}}%
  \includegraphics[height=3.34cm]{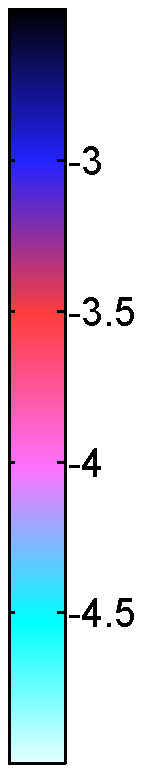}\put(0,35){\scriptsize \fontfamily{phv}\selectfont \rotatebox{90}{log g/cm\textsuperscript{\tiny 2}}}

  \caption{Column density of the central 12 $\times$ 12 arcsec$^2$ for NF, OF, OBBP, OB5, and OB10 (left-hand to right-hand side).  The arrows indicate the projected velocities of the stars.  The tick marks are at every 2$\times$10$^{17}$ cm, and the IRS~13E cluster is labelled.}
  \label{fi:SPHcoldens}
  \end{flushleft}
\end{figure*}

To briefly recapitulate the effect of the feedback mechanism on the hydrodynamic simulations, Fig.~\ref{fi:SPHcoldens} shows the column density at the present day for all models.  The most important point for this work is that the amount of material remaining in the simulation volume decreases as the strength of the feedback increases (left-hand to right-hand side).

\subsection{Radiative Transfer}\label{RT}

We solve the formal solution of radiative transfer along a 500$\times$500 grid of rays covering the central 15 $\times$ 15 arcsec$^2$ through the simulation volume for 0.3--12 keV [covering the \textit{Chandra} High Energy Transmission Grating (HETG) response function] at a resolution of 800 energy bins per dex.  This is more than ample for a majority of the energy range but is necessary to produce the correct shape around the Fe complex at $\sim$6.7~keV.  These pixel maps are folded through the \textit{Chandra} Advanced CCD Imaging Spectrometer-Spectroscopy (ACIS-S)/HETG zeroth-order response function to compare with the observed spectrum, and then through the \textit{Chandra} point spread function (PSF) [which is approximated as a 0.5 arcsec full width at half-maximum (FWHM) Gaussian] to compare with the observed image\footnote{We chose the order of folding through the response function first, summing over the 4--9 keV region second, and folding through the PSF third since it yields a significantly less intensive computation due to only PSF-folding one set of pixels.  We tested doing the PSF folding first (as is done on board \textit{Chandra}) over a subset of the full pixel and energy range, and the results are identical.}.
The response function for the model folding is generated in the analysis of the XVP observations via the standard \textsc{ciao} processing routines, as discussed in Section~\ref{Obs}.

The basis of the radiative transfer calculation (see \citealt{Russell13}; \citealt{RussellP16}, for more details) is the SPH visualization program \textsc{splash} \citep{Price07}.  The X-ray emissivities are from the \texttt{VVAPEC} model [a variation of the \texttt{APEC} model \citep{SmithP01}, which allows the abundance of each element from H to Zn to be set individually] using \texttt{AtomDB} \citep{FosterP12}
version 2.0.2, as implemented in \textsc{xspec} \citep{Arnaud96} version 12.0.9c.  The wind opacities are from \citet{VernerYakovlev95}\footnote{We tabulated the opacities in the manner first done by \citet{LeuteneggerP10} for the \texttt{windtabs} model, though here we modified the abundances to be for the WR stars in question.}, and the interstellar medium (ISM) opacities are from the \texttt{TBabs} model \citep*{WilmsAllenMcCray00}.  As the emissivities and wind opacities are metallicity dependent, we use three models to cover the range of WR spectral types in the SPH simulations: one for the WC stars, one for WN5-7, and one for WN8-9 and Ofpe/WN9.  Table \ref{RT:abu} lists the dominant elements.

Based on the column densities (Fig.~\ref{fi:SPHcoldens}) and wind opacities, the X-ray calculation over the simulation domain is well into the optically thin limit, particularly since only photons of energies $>$\,1\,keV are detected due to the high ISM absorption column.  As such, we use the aforementioned components to solve $I_E(x,y)=\int j_E(x,y)\,{\rm d}z$ over the simulation volume, where $I_E$ is the intensity at energy $E$, $(x,y)$ denotes a particular pixel, and $z$ is the direction towards the observer.  $j_E=n_{\rm e}n_{\rm i}\Lambda(E,T)$ is the emissivity of a parcel of gas with electron number density $n_{\rm e}$, ion number density $n_{\rm i}$, and temperature $T$.  The emission function $\Lambda(E,T)$ is from \texttt{VVAPEC}.

The only free parameter in the model is the ISM absorption column.  To determine $N_{\rm H}$, we upload our model spectra into \textsc{xspec} version 12.9.0n (using a custom-written model interface to import a text file directly into \textsc{xspec}) and fit the model to the data using the standard fitting procedure.  As will be shown (Section \ref{ISM}), the best-fitting value is $N_{\rm H}$\,=\,1.1$\times$10$^{23}$\,cm$^{-2}$.  We use this value to make all the X-ray images and spectra in this paper, thus allowing a direct comparison between all models.

\subsection{Observations}\label{Obs}

The XVP observations and data reduction are detailed in \citet[supplementary materials]{WangP13}.  Briefly, the \textit{Chandra} observations were taken with the HETG inserted in the light path and the ACIS-S at the focus.  Only the zeroth-order ACIS-S/HETG data are used in this work and were reduced via standard \textsc{ciao} processing routines (version 4.5; Calibration Database version 4.5.6).  In addition, the non-X-ray instrument background, estimated from the ACIS-S stowed data, was subtracted in the imaging analysis presented in this paper. For the spectral analysis, the local background subtraction method was used.

We perform the model-to-observation comparison with two X-ray data products of ACIS-S/HETG zeroth-order emission centred on \SAs: the 12 $\times$ 12 arcsec$^2$ image of 4--9~keV intensity and the spectrum from the 2--5 arcsec ring.  The former provides an upper limit to the colliding-wind emission this work attempts to model, while the latter provides a lower limit, as a result of the different methods of subtracting the X-ray background described here.

\begin{table}
  {\centering
  \caption{Mass fractions (per cent) of the three different abundance models for the X-ray emissivity and circumstellar absorption.}
  \label{RT:abu}
  \begin{tabular*}{\columnwidth}{l @{\extracolsep{\fill}} c c c c c}
    \hline\vspace{0.1em}\rule{0pt}{1em}
             & $X_\textrm{H}$        & $X_\textrm{He}$       & $X_\textrm{C}$          & $X_\textrm{N}$        & $X_\textrm{O}$
    \\ \hline\rule{0pt}{1.1em}
    \hspace{-0.2cm}WC8-9    & 0                     & 60                    & 31                      & 0                     & 7
    \\ \rule{0pt}{0em}
    \hspace{-0.2cm}WN5-7    & 0                     & 98.5                  & 0.029                   & 1                     & 0.018
    \\ \rule{0pt}{0em}
    \hspace{-0.2cm}WN8-9 \& Ofpe/WN9 & 11.5        & 82.4                  & 0.0124                  & 1.15                  & 0.066
    \\ \vspace{-1.2em}\\ \hline
  \end{tabular*}}
  \vspace{-0.8mm}
  \\ \textit{Notes.} WC8-9: \citet{Crowther07}; WN5-7: \citet*{Onifer08}; and WN8-9 and Ofpe/WN9: \citet*{HeraldHillierSchulteLadbeck01} obtained via the \textsc{CMFGEN} website.
\end{table}

\begin{figure}
  \includegraphics[width=\columnwidth]{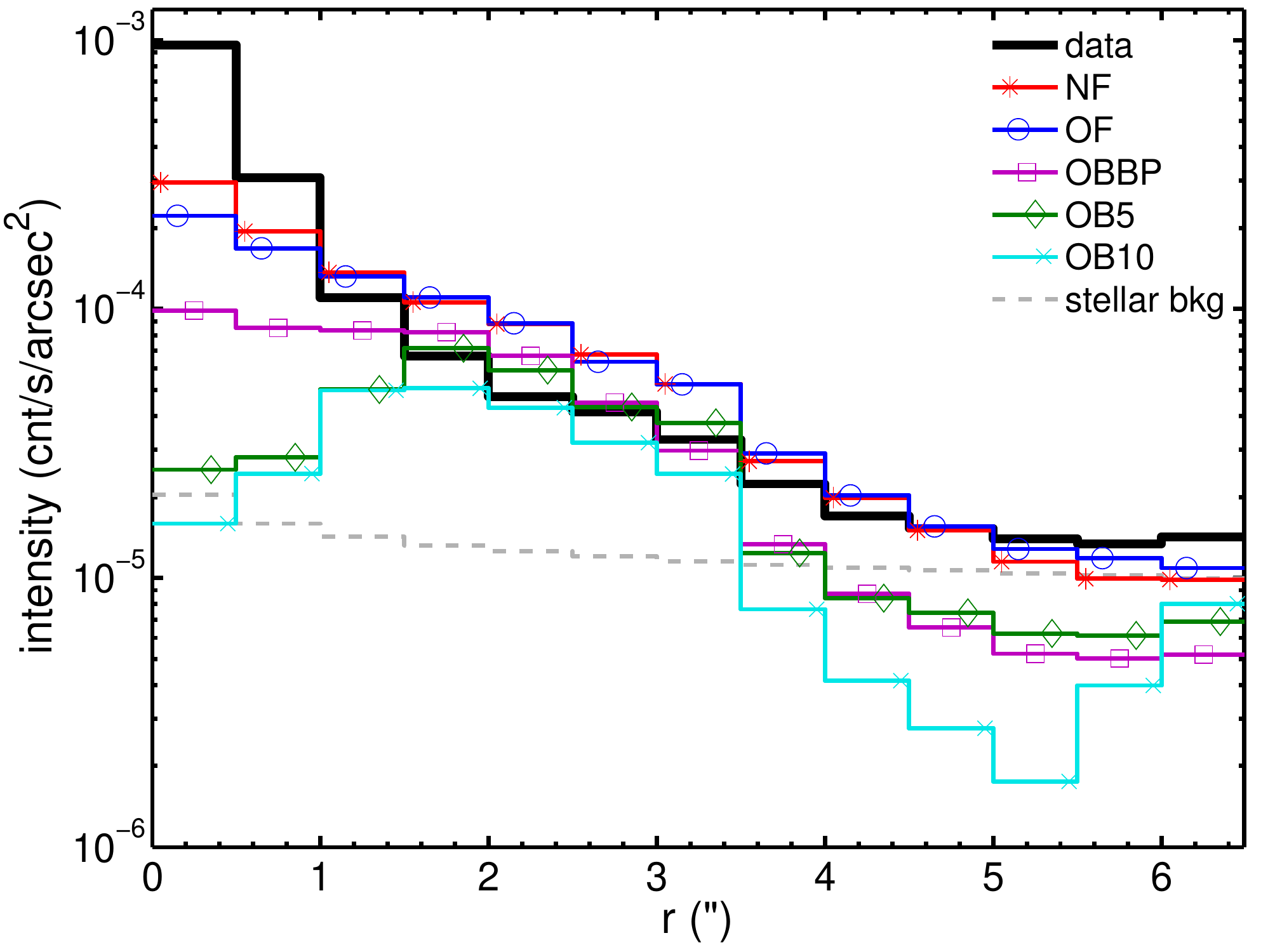}
  \caption{4--9 keV intensity of the observation and models as a function of radius from \SAs.  The background component due to the stellar population of CVs, which has been subtracted from the observation, is shown.}
  \label{fi:fr}
\end{figure}

As is typical in sources of extended X-ray emission, there is not a region void of source photons to use as an X-ray-background subtraction region; colliding-wind photons pervade the region of interest.  We therefore estimate the non-colliding-wind emission and subtract it from the 4--9~keV image.  The dominant component of the non-colliding-wind emission is thought to be unresolved cataclysmic variables (CVs).  We use the nuclear stellar cluster model of \citet[equations 17--19]{ChatzopoulosP15} to compute the radial stellar mass distribution and then the X-ray energy flux distribution using the ratio of the mass to the 2--8 keV luminosity given in \citet{GeP15}. We finally convert the flux into the 4--9 keV count intensity, using the conversion ratio of 1.3$\times$10$^{-10}$ (erg s$^{-1}$ cm$^{-2}$)/(count s$^{-1}$), obtained from the same intrinsic spectral model as used  in  \citeauthor{GeP15} but with the foreground absorption column density of \SAs. The final predicted radial intensity contribution from stars is shown as the  `stellar bkg' curve in Fig.~\ref{fi:fr}, which is subtracted from the 4--9 keV image.  The uncertainty in the predicted stellar contribution is probably dominated by the uncertainty ($\sim$21 per cent) in the extinction correction for the stellar near-infrared (near-IR) luminosity \citep{ChatzopoulosP15}.

\begin{figure*}
  \includegraphics[height=5cm]{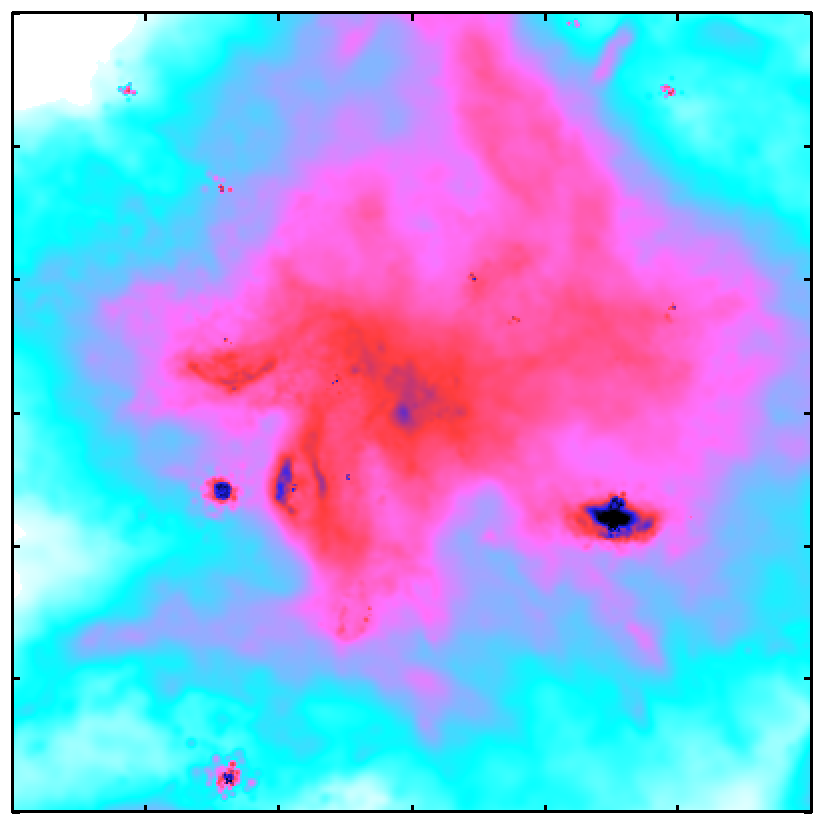}
  \put(-138,10){\scriptsize \fontfamily{phv}\selectfont \textbf{NF}}
  \put(-138,4){\scriptsize \fontfamily{phv}\selectfont \textbf{no PSF}}
  \put(-78,146){\vector(-1,0){63}}
  \put(-65,146){\vector(1,0){63}}
  \put(-76,144){\scriptsize \fontfamily{phv}\selectfont \textbf{12\arcsec}}
  \put(-148,67){\vector(0,-1){65}}
  \put(-148,75){\vector(0,1){65}}
  \put(-153,69){\scriptsize \fontfamily{phv}\selectfont \textbf{12\arcsec}}
  \includegraphics[height=5cm]{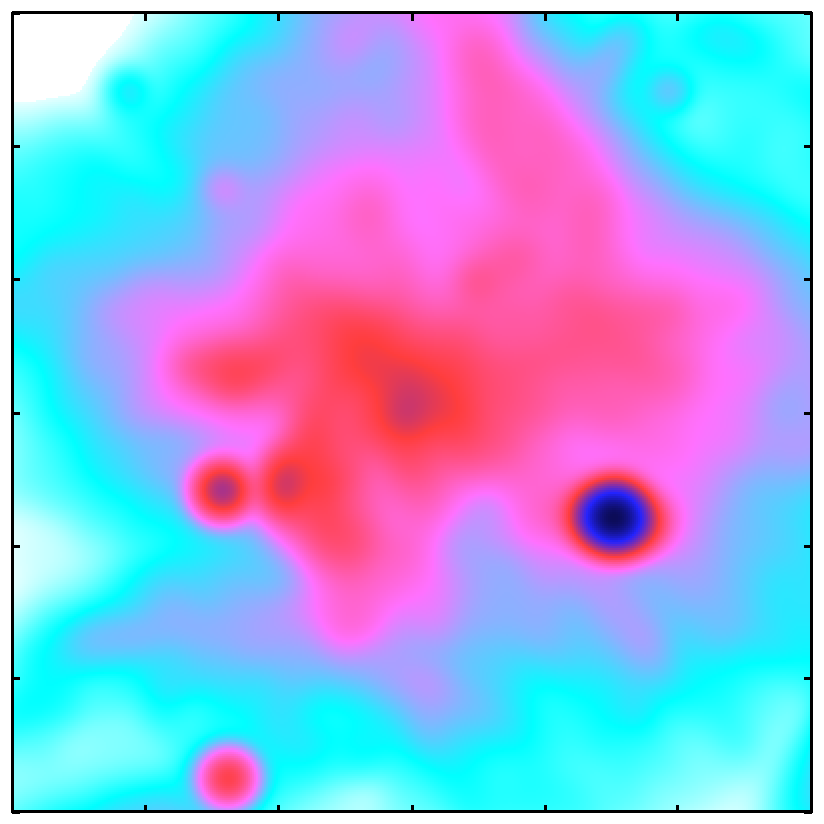}
  \put(-138,10){\scriptsize \fontfamily{phv}\selectfont \textbf{NF}}
  \put(-138,4){\scriptsize \fontfamily{phv}\selectfont \textbf{with PSF}}%
  \includegraphics[height=5cm]{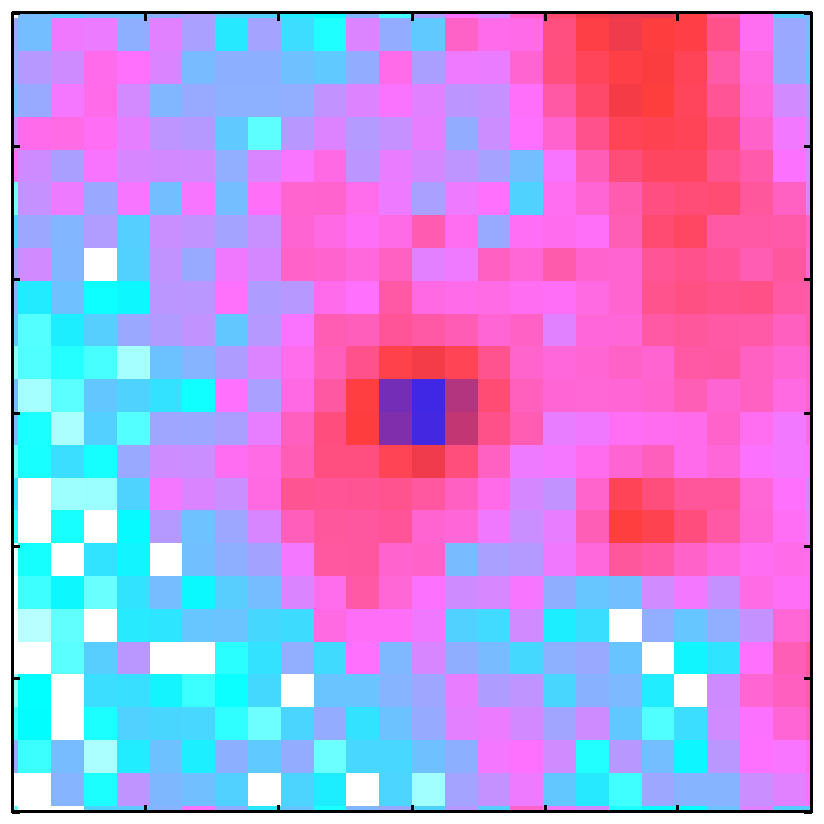}
  \put(-138,4){\scriptsize \fontfamily{phv}\selectfont \textbf{data}}
  \put(-40,128){\scriptsize \fontfamily{phv}\selectfont \textbf{PWN}}
  \put(-42,42){\scriptsize \fontfamily{phv}\selectfont \textbf{IRS~13E}}%
  \includegraphics[height=5cm]{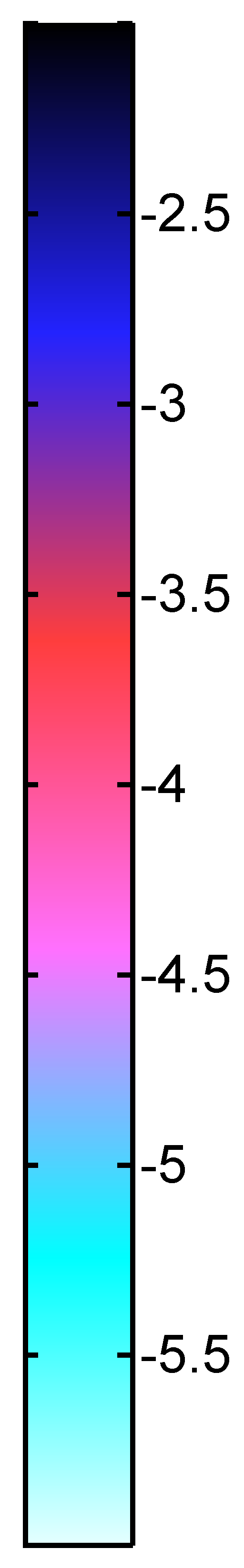}\put(3,49){\scriptsize \fontfamily{phv}\selectfont \rotatebox{90}{log cnt/s/arcsec\textsuperscript{\tiny 2}}}

  \vspace{-0.05cm}\caption{\textit{Chandra} 4--9 keV ACIS-S/HETG zeroth-order image (12 $\times$ 12 arcsec$^2$) of the control model NF (left-hand panel), model NF folded through the PSF (middle panel), and the data (right-hand panel).  The central pixels in the observation have a high intensity since they include the flaring from \SAs. The extended emission from the PWN, which is located just above the boundary of the plot, is labelled.}
  \label{fi:imC}
\end{figure*}

\begin{figure}
  \includegraphics[trim={7mm 4mm 1mm 1mm},clip,width=\columnwidth]{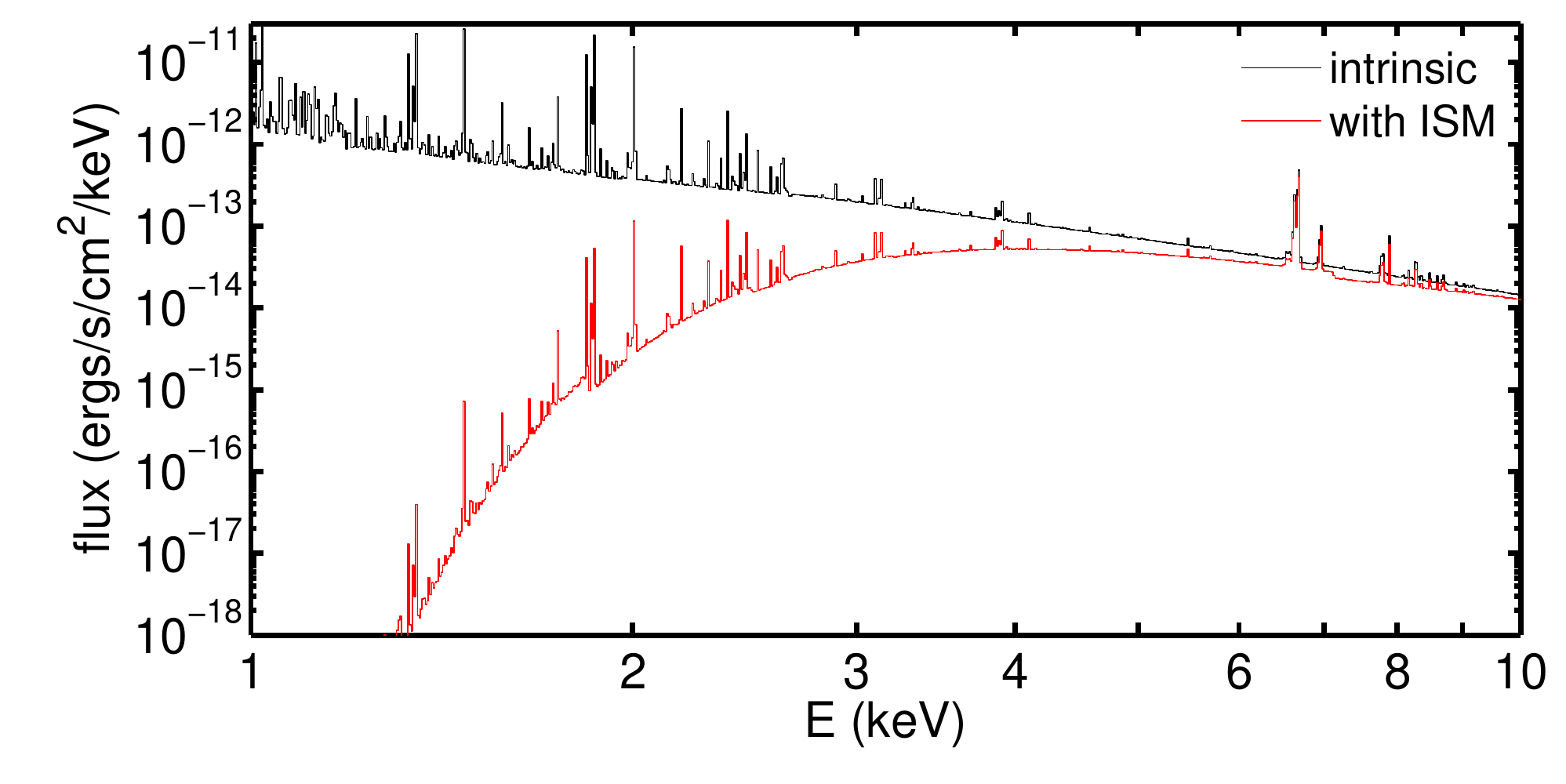}
  \vspace{-0.5cm}\caption{Control model NF spectra from the 2--5 arcsec ring centred on \SAs prior to folding through the response function.  The ISM strongly diminishes the X-ray flux below 3 keV.}
  \label{fi:spCnoresp}
\end{figure}

\begin{figure}
  \includegraphics[trim={7mm 4mm 1mm 1mm},clip,width=\columnwidth]{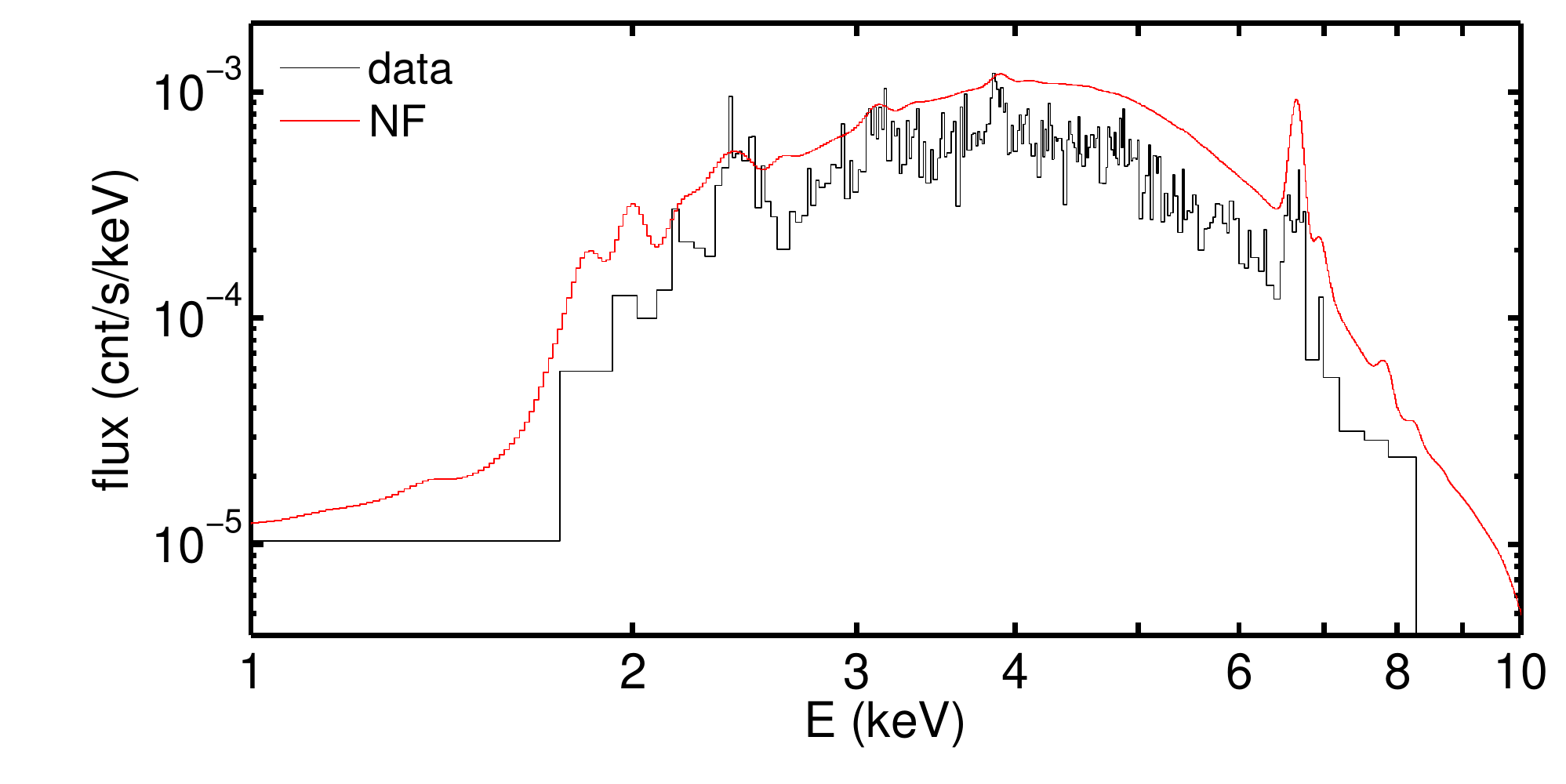}
  \vspace{-0.5cm}\caption{\textit{Chandra} ACIS-S/HETG zeroth-order spectra of the control model NF (red) and the data (black) from the 2--5 arcsec ring centred on \SAs, excluding the IRS~13E and PWN contributions.}
  \label{fi:spC}
\end{figure}

There are other X-ray-background sources, however, that are not accounted for in the 4--9 keV image, such as the X-ray emission far from \SAs, that is, outside the simulation volume of $r$ $<$ 12 arcsec.  This lack of accounting for the full X-ray background makes the image an upper limit.  On the other hand, the spectrum from the 2--5 arcsec ring has an available background-subtraction region, which \citet{WangP13} use as the ring of 6--18 arcsec.  Unfortunately, this region also includes colliding-wind emission, so the background subtraction is too strong and the spectrum is a lower limit.  As a result, the true discrepancy between the observation and model is somewhere between the model--image discrepancy and the model--spectrum discrepancy.

\begin{figure*}
  \begin{flushleft}

  \includegraphics[height=3.37cm]{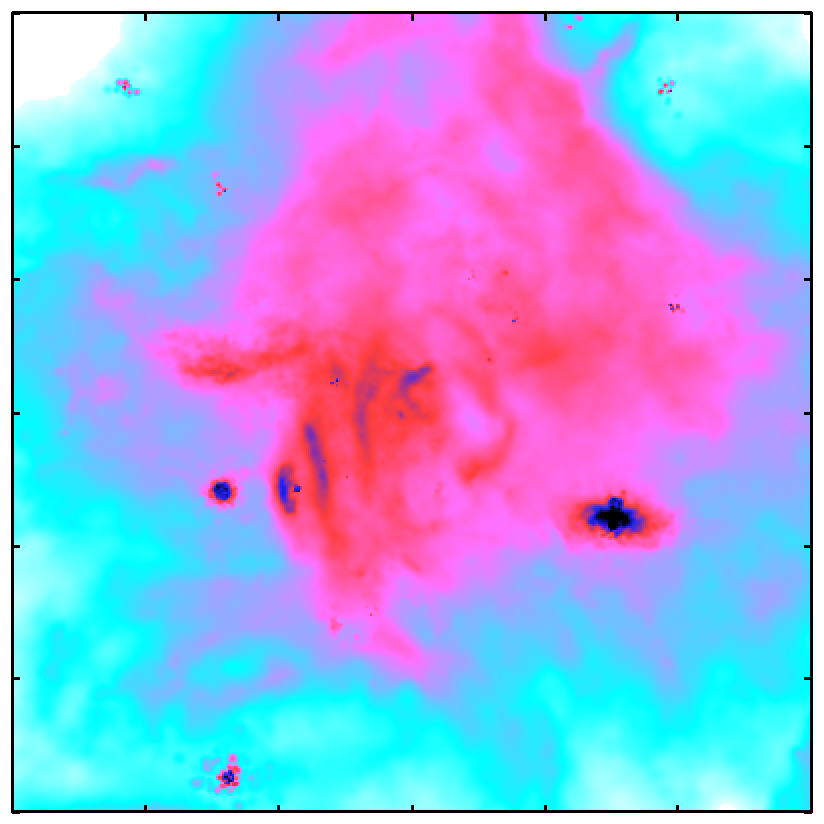}
  \put(-11,10){\scriptsize \fontfamily{phv}\selectfont \textbf{OF}}
  \put(-23.7,4){\scriptsize \fontfamily{phv}\selectfont \textbf{no PSF}}%
  \includegraphics[height=3.37cm]{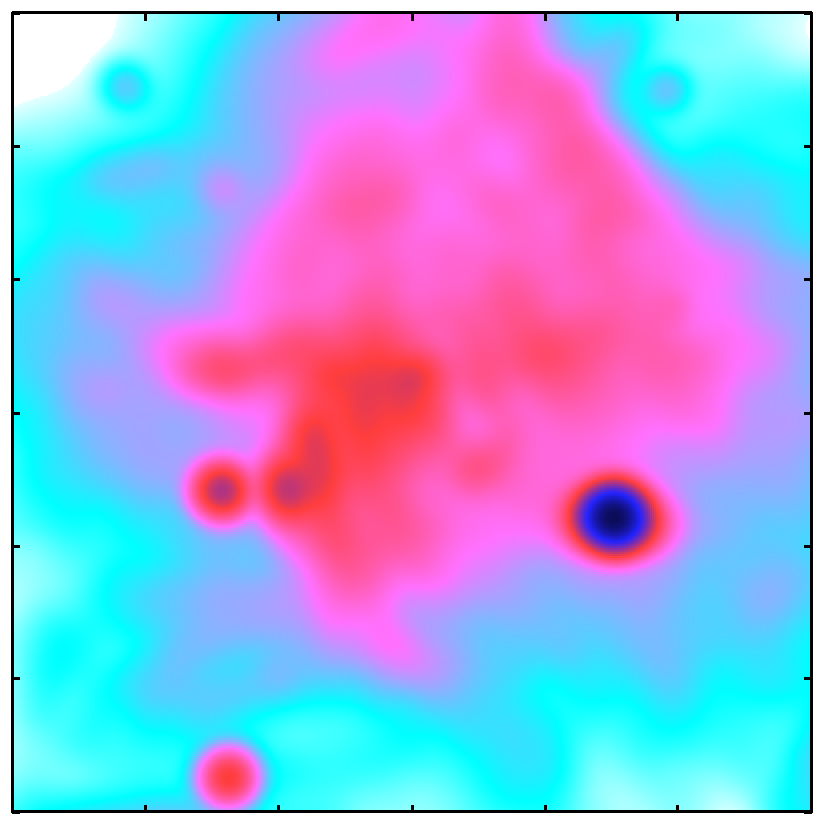}
  \put(-11,10){\scriptsize \fontfamily{phv}\selectfont \textbf{OF}}
  \put(-28.4,4){\scriptsize \fontfamily{phv}\selectfont \textbf{with PSF}}%
  \includegraphics[height=3.37cm]{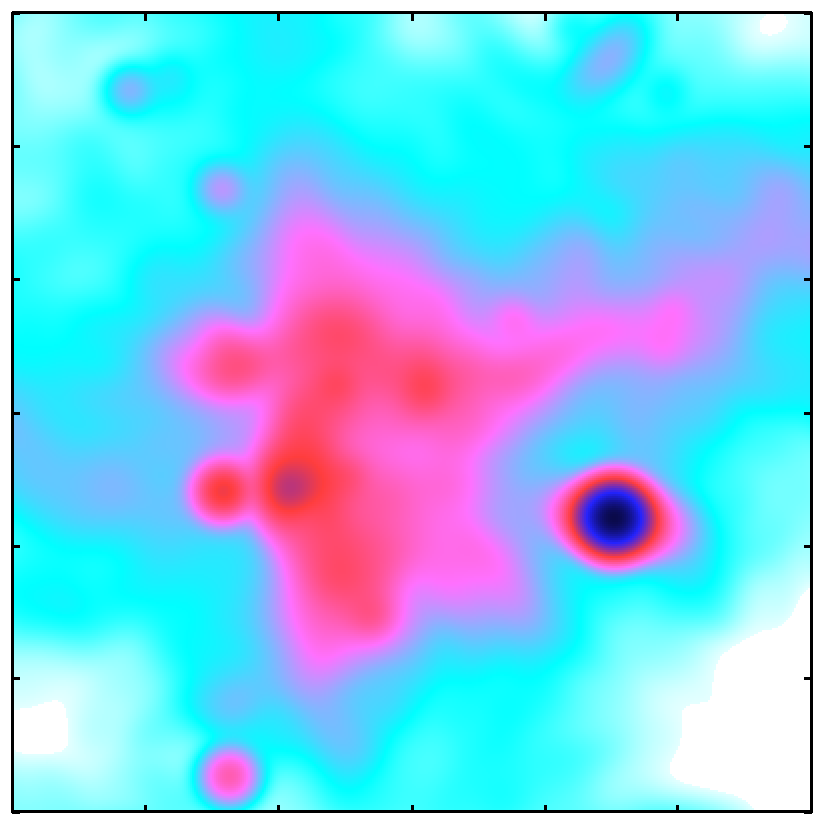}
  \put(-20,10){\scriptsize \fontfamily{phv}\selectfont \textbf{OBBP}}
  \put(-28.4,4){\scriptsize \fontfamily{phv}\selectfont \textbf{with PSF}}
  \put(-54,100){\vector(-1,0){40}}
  \put(-42,100){\vector(1,0){40}}
  \put(-53,98){\scriptsize \fontfamily{phv}\selectfont \textbf{12\arcsec}}%
  \includegraphics[height=3.37cm]{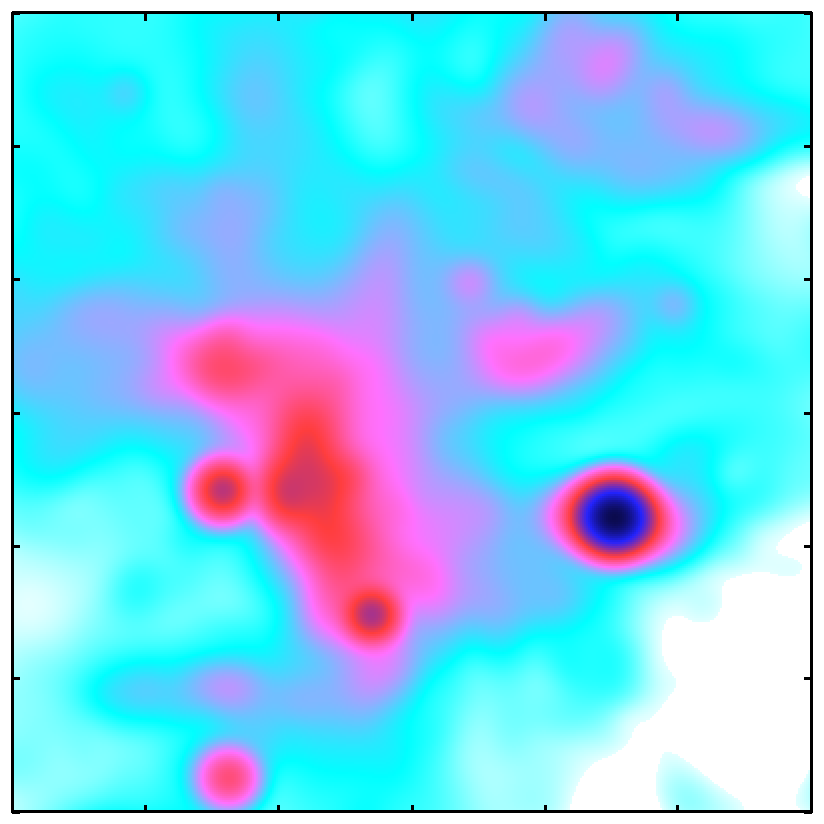}
  \put(-15.4,10){\scriptsize \fontfamily{phv}\selectfont \textbf{OB5}}
  \put(-28.4,4){\scriptsize \fontfamily{phv}\selectfont \textbf{with PSF}}%
  \includegraphics[height=3.37cm]{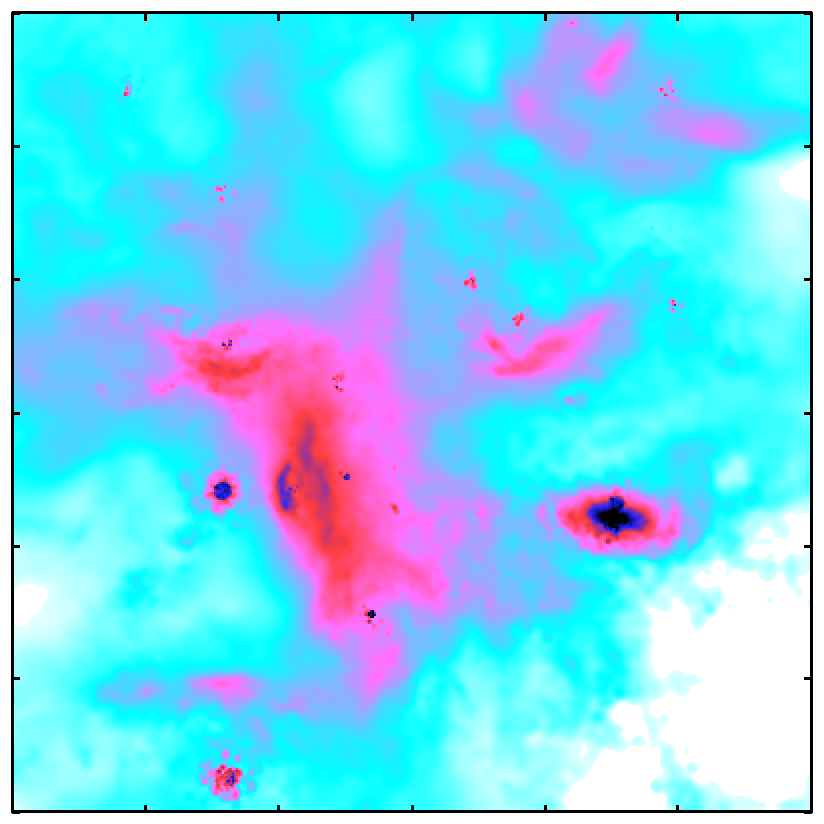}
  \put(-15.4,10){\scriptsize \fontfamily{phv}\selectfont \textbf{OB5}}
  \put(-23.7,4){\scriptsize \fontfamily{phv}\selectfont \textbf{no PSF}}%
  \includegraphics[height=3.37cm]{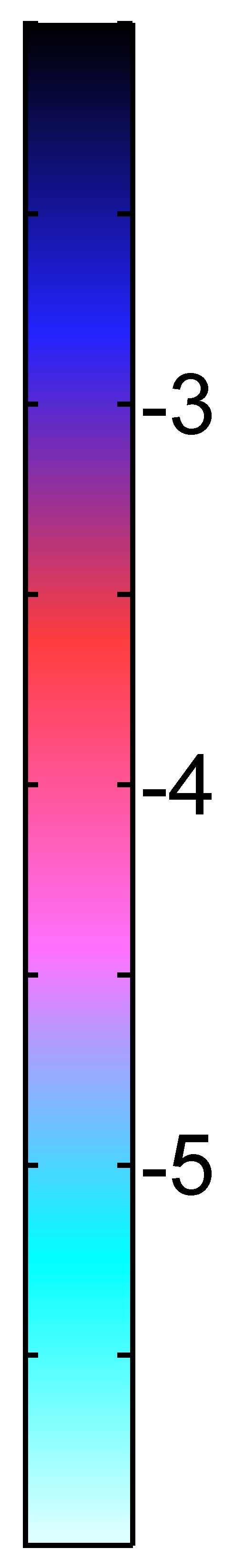}
  \put(1,26){\scriptsize \fontfamily{phv}\selectfont \rotatebox{90}{log cnt/s/arcsec\textsuperscript{\tiny 2}}}

  \vspace{-0.04cm}

  \includegraphics[height=3.37cm]{GCImage_g3_4to9_new179_8xNO7p5_NSTC_r100.png}
  \put(-11.4,10){\scriptsize \fontfamily{phv}\selectfont \textbf{NF}}
  \put(-23.7,4){\scriptsize \fontfamily{phv}\selectfont \textbf{no PSF}}
  \put(-35,28){\scriptsize \fontfamily{phv}\selectfont \textbf{IRS~13E}}%
  \includegraphics[height=3.37cm]{GCImage_g3_4to9_new179_8xNO7p5_PSF_NSTC_r100.png}
  \put(-11.4,10){\scriptsize \fontfamily{phv}\selectfont \textbf{NF}}
  \put(-28.4,4){\scriptsize \fontfamily{phv}\selectfont \textbf{with PSF}}%
  \includegraphics[height=3.37cm]{GCImage_Data_4to9keV_r100.png}
  \put(-15.6,4){\scriptsize \fontfamily{phv}\selectfont \textbf{data}}%
  \includegraphics[height=3.37cm]{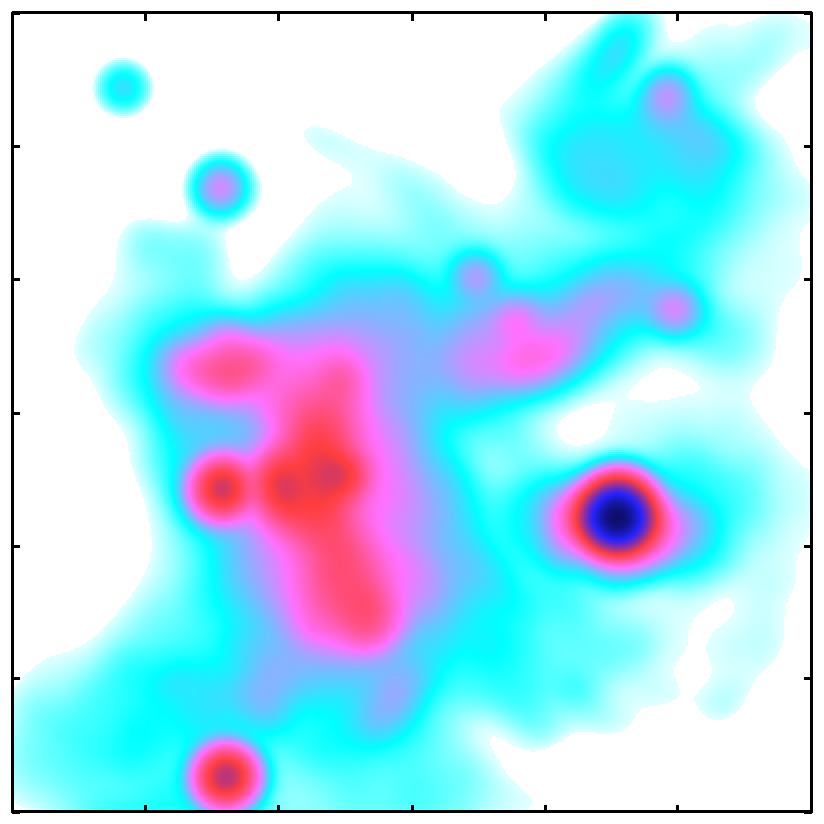}
  \put(-18.7,10){\scriptsize \fontfamily{phv}\selectfont \textbf{OB10}}
  \put(-28.4,4){\scriptsize \fontfamily{phv}\selectfont \textbf{with PSF}}%
  \includegraphics[height=3.37cm]{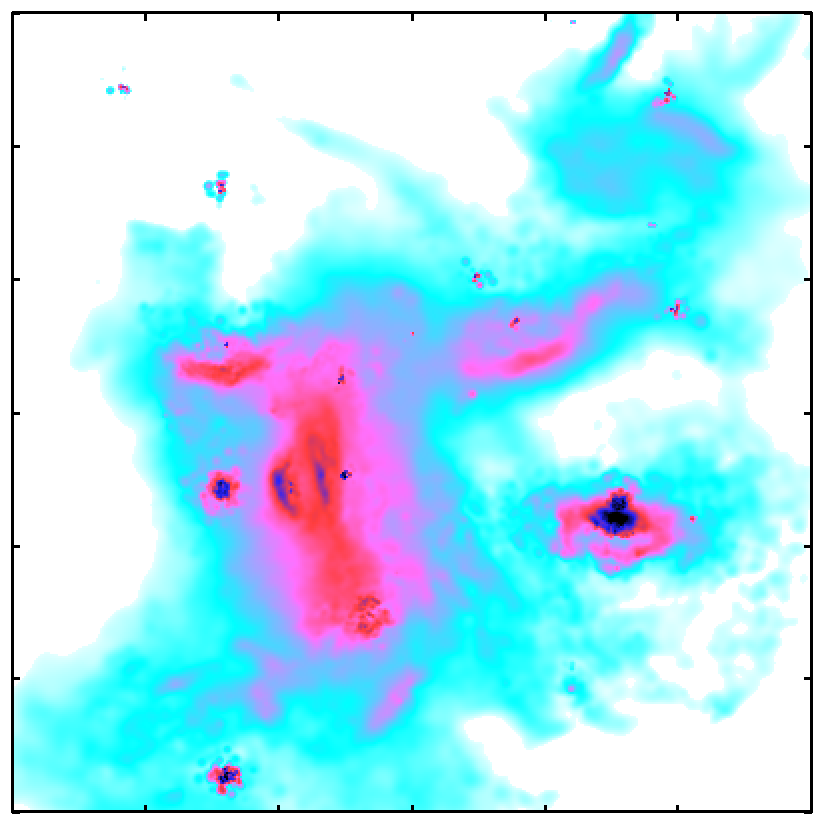}
  \put(-18.7,10){\scriptsize \fontfamily{phv}\selectfont \textbf{OB10}}
  \put(-23.7,4){\scriptsize \fontfamily{phv}\selectfont \textbf{no PSF}}%
  \includegraphics[height=3.37cm]{GCImage_ColorBar_4to9keV_BigFont_NSTC.png}
  \put(1,26){\scriptsize \fontfamily{phv}\selectfont \rotatebox{90}{log cnt/s/arcsec\textsuperscript{\tiny 2}}}
  \caption{Same as Fig.~\ref{fi:imC}, but for all models.  The data are in the bottom middle panel.  For the models, the first and last columns are not folded through the PSF, while the three middle columns are.  Going clockwise in decreasing X-ray flux, the bottom left-hand panels are the control model NF, the top left-hand panels are OF, the top middle panel is OBBP, the top right-hand panels are OB5, and the bottom right-hand panels are OB10.  The NF and OF models are very similar, while the outburst models show decreasing X-ray flux as the feedback increases.}
  \label{fi:imFB}
  \end{flushleft}
\end{figure*}

\begin{figure}
  \includegraphics[width=\columnwidth]{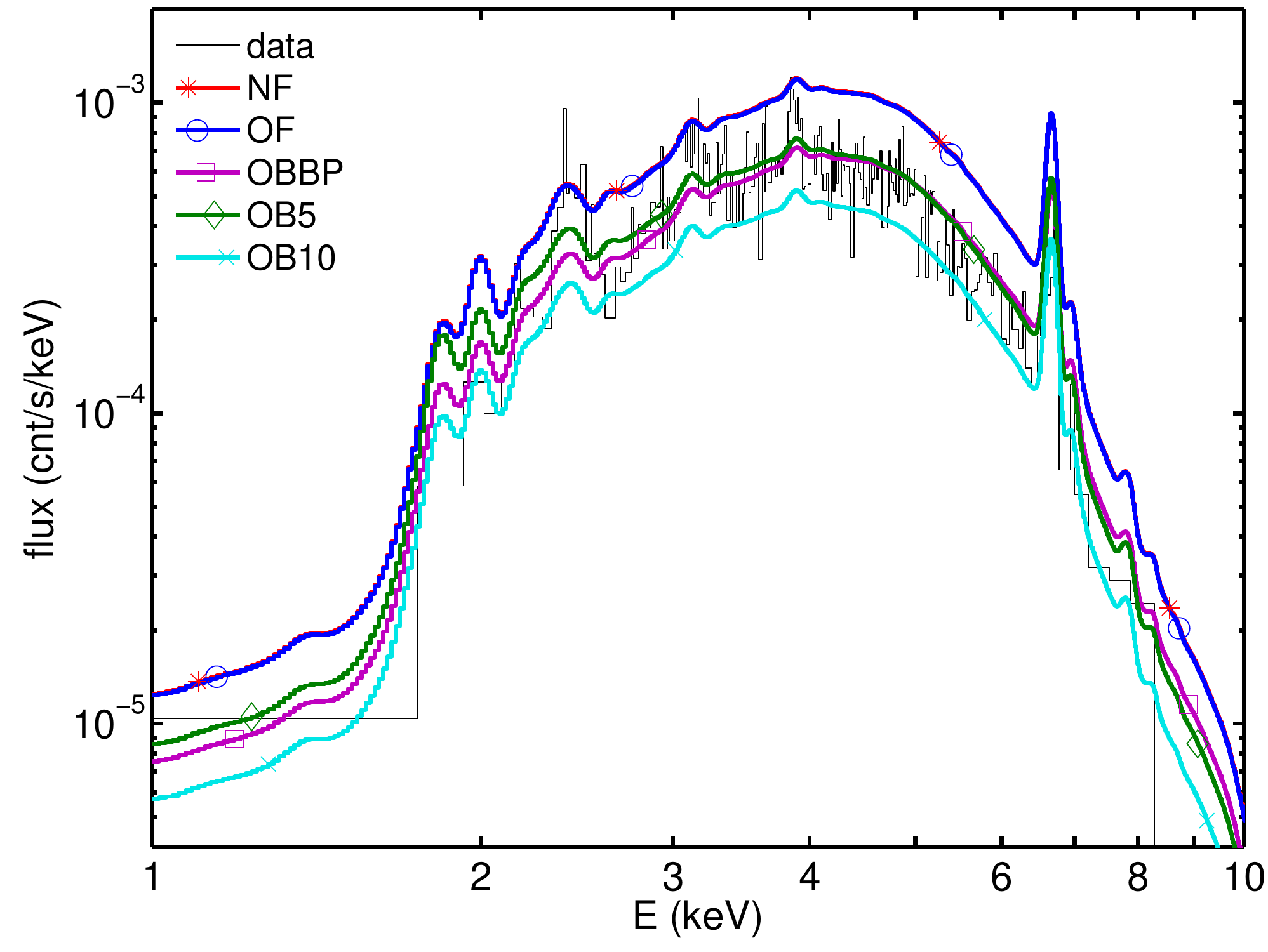}
  \caption{Same as Fig.~\ref{fi:spC}, but for all models.  As expected, the level of X-ray flux follows the trend of Fig.~\ref{fi:imFB}.}
  \label{fi:spFB}
\end{figure}

It is also worth noting that the SMBH point-source emission and the pulsar wind nebula (PWN) extended emission are not modelled, so these regions should not be compared in the images.  The spectra (source and background) include neither the PWN region nor point sources,\footnote{See fig.~S.1 of the supplementary material of \citet{WangP13} for the extraction regions of the 2--5 arcsec (source) and 6--18 arcsec (background) rings.} such as IRS~13E, which makes the comparison of the diffuse model emission with the observations as accurate as possible.

\begin{table}
  \centering
  \caption{Model-to-data ratio of the 4--9 keV flux from the 2--5 arcsec ring, excluding IRS~13E and the PWN.  The differences in the image and spectral results account for the different X-ray-background estimations for each observable, resulting in the observed image flux being an upper limit and the observed spectral flux being a lower limit (Section \ref{Obs}).}
  \label{ta:ModelVsData}
  \begin{tabular*}{\columnwidth}{l @{\extracolsep{\fill}} c c c}
  \hline\vspace{0.1em}\rule{0pt}{1em}
    \hspace{-0.2cm}Model & Image & Spectra & Mean \\ \hline\rule{0pt}{1.1em} 
    \hspace{-0.2cm}NF    & 1.50  & 1.97    & 1.73 \\ \rule{0pt}{0em} 
    \hspace{-0.2cm}OF    & 1.50  & 1.96    & 1.72 \\ \rule{0pt}{0em} 
    \hspace{-0.2cm}OBBP  & 0.92  & 1.21    & 1.07 \\ \rule{0pt}{0em} 
    \hspace{-0.2cm}OB5   & 0.92  & 1.20    & 1.06 \\ \rule{0pt}{0em} 
    \hspace{-0.2cm}OB10  & 0.62  & 0.81    & 0.71 \\ \vspace{-1.2em}\\ \hline
  \end{tabular*}
\end{table}

Due to folding the model spectra through the \textit{Chandra} response function, smoothing the image with the PSF, and accounting for the distance to the Galactic Centre, the comparisons of the model images and spectra with the data (subject to the observational uncertainties discussed above) are exact.  There are no scaling or normalization factors that can arbitrarily increase or decrease the model X-ray levels, so the comparisons provide stringent tests of the hydrodynamic simulations for determining the density and temperature structure of the stellar-wind-ejected material around \SAs.

\section{Results}\label{R}

\subsection{Control model -- no feedback}

To provide more information about the X-ray calculation, we first present the control model (NF) in more detail than the others.
Fig.~\ref{fi:imC} shows the 4--9 keV flux image from NF compared with the data, both before and after accounting for the \textit{Chandra} PSF.
The overall agreement is good.  The model flux in the range of $1.5<r<4$ arcsec is about a factor of $\sim$2 higher than the observation, that in the range of $4.5<r<5.5$ arcsec agrees well, and from $r>6$ arcsec the model flux is too low.  Summed over the 2--5 arcsec ring, excluding IRS~13E and the PWN, the model is 1.50 times higher than the data (see Table \ref{ta:ModelVsData}).  The largest discrepancy is IRS~13E, where the model flux is approximately seven times higher than the observation, and is sensitive to the assumed specific wind properties of the two WRs in this compact cluster.

Fig.~\ref{fi:spCnoresp} shows the model spectrum for the 2--5 arcsec ring (excluding IRS~13E and the PWN) prior to the response function folding, both before and after accounting for the strong influence of the ISM absorption, while Fig.~\ref{fi:spC} compares the observed spectrum with the model one after being folded through the ACIS-S/HETG zeroth-order response function.
The model flux in the 4--9 keV band is 1.97 times higher than the observed one (see Table \ref{ta:ModelVsData}).  Therefore, a reasonable estimate of the actual discrepancy (taken to be the mean of the model-image deviation and the model-spectrum deviation) in the 2--5 arcsec region is a factor of $\sim$1.7.

A noteworthy achievement of the model is the agreement in the spectral shape (Fig.~\ref{fi:spC}).  This provides strong evidence that the WR wind--wind collisions are the dominant source of X-ray emission around \SAs.  The temperature of the gas around \SAs is naturally explained by the WR shocked material, validating the wind speeds of the WR stars used in the simulations.

\subsection{Feedback Models}

Figs \ref{fi:imFB} and \ref{fi:spFB} show the 4--9 keV images and the 2--5 arcsec ring spectra for all models (with and without feedback).  The OF results are very similar to the NF results, indicating that this weak feedback has only a small effect on the X-ray emission.
For the outburst models, the strong SMBH wind evacuates material from the simulation volume, including the hot ambient medium, which has built up over time out of the stellar winds, so the OBBP, OB5, and OB10 models have increasingly less X-ray emission.
The similarity in shapes of the outburst-model spectra to the control and outflow models is expected because  only WR-ejected material is creating the thermal X-ray emission; the faster moving outburst material, which would produce harder spectra, left the $r<12$ arcsec simulation volume since the outbursts ceased 100 yr ago.

To compare the model and data images more quantitatively, Fig.~\ref{fi:fr} shows the 4--9 keV intensity as a function of projected radius from \SAs.  As was done for the spectral comparison, the emission from IRS~13E and the PWN is excluded for both the data and model intensity profiles.  Since the model flux does not include any point-source contribution from \SAs while the data do, the model-data comparison is only valid on scales greater than the $\sim$0.5 arcsec FWHM PSF.

The differences in NF and OF are clearer in Fig.~\ref{fi:fr} than in the images; the outflow clears out some material just around \SAs, making the X-ray emission lower for $<$1 arcsec.  Their behaviour is similar beyond this radius, thereby showing the limited extent of the outflow feedback mechanism.  Like the images, the intensity profiles of the outburst models show the trend of higher feedback leading to weaker X-ray emission, particularly within 1 arcsec of \SAs.  The best-matched region around 2 arcsec is also the location of the majority of the WR stars.  This indicates that the cessation time of the outburst affects the X-ray emission, so perhaps these outburst models could reproduce the observations if they were ended sooner.  Furthermore, since the best-matched region for all models is the WR-rich region of $\sim$2 arcsec, this provides more evidence that the dominant contribution to the thermal X-ray emission around \SAs is the WR-wind collisions.

The model that best agrees with the observation over 2--5 arcsec (excluding IRS~13E and the PWN) is the intermediate-strength feedback model OB5 (see Table \ref{ta:ModelVsData}).  The model-to-data flux ratio over 4--9~keV is 0.92 and 1.20 in the image and spectra comparison, respectively, resulting in a mean discrepancy of the model having $\sim$6 per cent higher flux than the observation.

\section{Discussion}\label{D}

\subsection{ISM absorption column}\label{ISM}

Table \ref{ta:nH} shows the results of the ISM-absorption fitting, namely the ratio of the model flux to the observed flux\footnote{Note that these values differ from the `spectra' values in Table \ref{ta:ModelVsData} since these values are based on fitting the whole spectrum, while Table \ref{ta:ModelVsData} uses only from 4 to 9 keV.}, the best-fitting $N_{\rm H}$, and the 90 per cent statistical confidence intervals as the errors.  The lowest $\chi^2_{\rm red}$ is from OB10, indicating that it has the closest shape to the observation; however, the flux of OB5 is much closer to the observation (see also Table~\ref{ta:ModelVsData}), while the spectral shape is only slightly worse, so OB5 is the best-fitting model.
Both these spectra have best-fitting $N_{\rm H}$ values of 1.1$\times$10$^{23}$\,cm$^{-2}$, which is lower than the value of $N_{\rm H}=1.66_{-0.25}^{+0.28}\times10^{23}$\,cm$^{-2}$ obtained by fitting the central \SAs source \citep{WangP13}.  One potential explanation is regarding the background subtraction of the 2--5 arcsec ring.  Its background region of 6--18 arcsec does not completely subtract off the CV-dominated component of the old stellar population \citep{GeP15} as this component falls off radially (see `stellar bkg' of Fig.~\ref{fi:fr}).  Since this stellar component has a soft spectrum \citep{GeP15,XuWangLi16}, fitting a spectrum without this component completely subtracted off would yield a lower $N_{\rm H}$ than otherwise expected.
Still, it is encouraging that there is moderate agreement between the $N_{\rm H}$ measurements of the two different regions, one solely including \SAs, and the other completely excluding \SAs.

\begin{table}
  \centering
  \caption{Results of the ISM absorption fitting showing the inverse norm [i.e.\ (model flux)/(observed flux)], $N_{\rm H}$ with 90 per cent confidence interval errors, and $\chi^2_{\rm red}$ for each model.  Each fit has 211 degrees of freedom.}
  \label{ta:nH}
  \begin{tabular*}{\columnwidth}{l @{\extracolsep{\fill}} c c c}
    \hline\vspace{0.3em}\rule{0pt}{1.2em}
    Model & Norm$^{-1}$ & $N_{\rm H}$ (10$^{22}$\,cm$^{-2}$) & $\chi^2_{\rm red}$ \\ \hline\rule{0pt}{1.3em} 
    NF    & 1.95        & 10.28$^{+0.29}_{-0.28}$            & 1.36 \\ \rule{0pt}{1.4em} 
    OF    & 1.94        & 10.26$^{+0.29}_{-0.28}$            & 1.36 \\ \rule{0pt}{1.4em} 
    OBBP  & 1.20        & 10.16$^{+0.29}_{-0.27}$            & 1.38 \\ \rule{0pt}{1.4em} 
    OB5   & 1.17        & 10.92$^{+0.30}_{-0.28}$            & 1.31 \\ \rule{0pt}{1.4em} 
    OB10  & 0.79        & 10.73$^{+0.31}_{-0.29}$            & 1.24 \\ \vspace{-0.8em}\\ \hline
  \end{tabular*}
\end{table}

\subsection{IRS~13E}

The largest source of discrepancy in the images for all models is the overabundance of X-rays coming from the IRS~13E cluster.  The model 4--9\,keV flux within a 1-arcsec circle centred on IRS~13E is approximately 7 times higher than the same region in the observation.  The pre-PSF model images (Figs \ref{fi:SPH}, \ref{fi:imC}, and \ref{fi:imFB}) show that the region between the two WRs in this cluster, which is where the winds from these two stars collide head-on, is very X-ray bright. There is no outside factor, such as the remnant of another WR passing nearby, that influences the emission.  The type of feedback also does not affect the IRS~13E emission; even with much wind material cleared out of the simulation volume in OB10, the 100 yr from when the outburst was shut off to the present day is enough time for the wind--wind collision between the stars to build back up and produce the same X-ray flux (see Fig.~\ref{fi:imFB}).

As such, the only way to decrease the IRS~13E X-ray flux is to make one or both of the winds weaker and/or increase the separation of the stars (without changing their locations on the sky).  Since these thermal X-rays are produced by collisions (a density-squared process), a factor of $\sim$2.5 reduction in one of their mass-loss rates will naively achieve the desired emission reduction of a factor of $\sim$7, though the changing wind--wind collision geometry \citep*[e.g.][]{CantoRagaWilkin96} makes this determination slightly more involved. Alternatively, since these are adiabatic shocks whose luminosity scales inversely proportional to the distance to the shock \citep*{LuoMcCrayMacLow90,StevensBlondinPollock92}, their separation needs to increase by a factor of $\sim$7 to agree with the models.  A more moderate adjustment of one or both mass-loss rates and the separation would also work.  To constrain the velocities of these winds, the model and observed spectra of these regions should be compared, which will be done in future work.

There are lower mass stars in IRS~13E, most notably a B star, that are not in these simulations.  However, their inclusion would seemingly increase only the X-ray emission since the WR winds can only shock closer to their stars if, say, the B star wind is in between the two WRs.  This should only be a minute increase, though, since the momentum of a typical B-star wind is orders of magnitudes lower than a WR wind.

It should also be mentioned that a different spectral classification of the two WR stars might also decrease the X-ray emission.  The shocked WC winds produce much more X-ray emission per unit mass than shocked WN winds, so if the WC star in IRS~13E is misclassified, then the model emission would be more consistent with the observations.  This misclassification, however, seems unlikely \citep{MartinsP07}, and, by extension, suggests that the WC wind is the wind in this cluster that should be weakened.

\subsection{Overall diffuse emission}

The diffuse emission surrounding \SAs (12 $\times$ 12 arcsec$^2$, excluding IRS~13E and the PWN) compares well with the observations.  This is particularly encouraging since the dominant diagnostics used to create the hydrodynamic models -- orbits, mass-loss rates, and wind speeds -- are from IR diagnostics.  The lone X-ray diagnostic used is in relation to the SMBH feedback, not from the thermal X-ray emission.  Therefore, it is noteworthy that hydrodynamic models constructed without the use of thermal X-ray constraints well reproduce the thermal X-ray properties of the Galactic Centre.

The modelling indicates that the wind speeds are well constrained due to the spectral matching.  While the ISM absorption column affects the soft portion of the X-ray spectrum, the portion $>$4 keV is unaffected by the ISM, so the good match between the models and the observations in this high energy range can be obtained only by proper wind speeds.  For the mass-loss rates, there is somewhat of a degeneracy with the SMBH feedback strength.  Either higher mass-loss rates and a stronger SMBH feedback mechanism, or lower mass-loss rates and a weaker SMBH feedback, could potentially reproduce the data just as well as the current OB5 model.

For all models, as the images of Fig. \ref{fi:imFB} hint at and the intensity versus radius plots of Fig. \ref{fi:fr} show more explicitly, the X-ray emission falls off too fast beyond $r=4$ arcsec. One possible solution is to include two structures that tend to confine the hot gas leaving the central region in certain directions: the `mini-spiral' and the circumnuclear disc \citep[e.g.][]{GenzelEisenhauerGillessen10}.
The `mini-spiral' is formed by streamers of ionized gas that occupy the region in question, while the circumnuclear disc encompasses it at $\sim$1.5 pc.  Both gas components could partially prevent the escape of hot gas.
As these features would influence only the X-ray emission in the outer regions, where large discrepancies between the simulations and the observations are seen, including the `mini-spiral' and circumnuclear disc should be explored in future work.

Incorporating other known elements could also change the X-ray emission.  The O stars are more numerous than the WRs, and while their winds are weaker than WR winds, their inclusion should cause the WR winds to shock closer to their originating stars.  This will increase the X-ray emission coming from these adiabatic WR-wind shocks.  Importantly, this alteration of the WR-wind shocks will not alter the spectral shape since the pre-shock speeds will be the same.  The O-star shocks should also produce some of their own X-ray emission, though much less than the WRs due to their weaker winds and lower emissivities.
To a lesser degree, incorporating lower mass stars with weaker winds (such as the S stars modelled in \citealt{LutzgendorfP16}) will also achieve the same effect.
Binaries are another potential source of additional X-ray emission.  Three WRs have been identified as binaries so far, while $\sim$30 per cent of the OB/WR stars are thought to be spectroscopic binaries \citep{PfuhlP14}. The closest of these binaries will produce radiative shocks, which cool quickly and produce little X-ray emission, but more widely separated binaries can contribute to the X-ray emission around \SAs.

Finally, there may be some potential effect due to non-equilibrium ionization of the hot gas. Intuitively, we do not expect this effect to be important enough to affect the results reported here because of the high density of the hot gas (hence, generally small ionization equilibrium time-scales). But this effect deserves a careful study in the future. Because of the substantial reduction of the effective area with the insertion of the HETG, the counting statistics of the XVP zeroth-order data used here are poor at low photon energies, where the relevant He-like-triplet emission lines [e.g. S \textsc{xv} ($\sim$2.4 keV), Ar \textsc{xvii} ($\sim$3.1 keV), and Ca \textsc{xix} ($\sim$3.9 keV)] are located. The future inclusion of observations without the grating will help to tighten the constraints on the emission lines and hence the potential non-equilibrium ionization effect.

\subsection{Clumps}

The column density plot (Fig.~\ref{fi:SPHcoldens}) shows the presence of many clumps that form in the hydrodynamic simulations. These are from collisions of winds with slow terminal speeds, which therefore cool and condense quickly \citep{CuadraNayakshinMartins08}.  However, the recent analytical study of \citet{CalderonP16} shows that the amount of clumps is probably overestimated in the simulations.  Fortunately, the clumps are cold and do not emit X-rays, and the soft X-rays they obscure are heavily obscured by the ISM also. Therefore, the uncertainty in the amount of clumps has a minimal impact on this work.

\section{Conclusions}\label{C}

We compute the thermal X-ray properties of the Galactic Centre from hydrodynamic simulations of the 30 WR stars orbiting  within 12 arcsec of \SAs \citep{CuadraNayakshinMartins08,CuadraNayakshinWang15}.  These simulations use different feedback models from the SMBH at its centre.  The \textit{Chandra} XVP observations \citep{WangP13} provide an anchor point for these simulations, so we compare the observed 12 $\times$ 12 arcsec$^2$ 4--9 keV image and the 2--5 arcsec ring spectrum with the same observables as synthesized from the models.  Remarkably, the shape of the model spectra, regardless of the type of feedback, agrees very well with the data.  This indicates that the hot gas around \SAs is primarily from shocked WR wind material, and that the velocities of these winds are well constrained.  The ISM absorption column from fitting this diffuse emission broadly agrees with the absorbing column determined from fitting the point-source emission of \SAs \citep{WangP13}.  The X-ray flux strongly depends on the feedback mechanism; greater SMBH outflows clear out more WR-ejected material around \SAs, thus decreasing the model X-ray emission.
Over 4--9~keV in energy and 2--5 arcsec in projected distance from \SAs (excluding IRS~13E and the nearby PWN), the X-ray emission from all models is within a factor of 2 of the observations, with the best model agreeing to within 10 per cent;
this is the intermediate-strength feedback model OB5,
which has an SMBH outburst of $\dot{M}_{\rm out}=10^{-4}$\,$M_\odot$\,yr$^{-1}$, $v$\,=\,5000\,km\,s$^{-1}$, and occurring from 400 to 100\,yr ago.
Therefore, this work shows that the SMBH outburst is required for fitting the X-ray data, and, by extension, that the outburst still affects the current X-ray emission surrounding \SAs, even though it ended 100\,yr ago.

Future work should address the completeness of the hydrodynamic simulations by adding other sources of gas in the simulation volume, such as the O stars and binaries that are located in the region currently modelled, and the `mini-spiral' and circumnuclear disc that are farther out and could prevent the hot gas from escaping the central region as easily.  Both these effects could increase the overall X-ray emission, thereby requiring a reduction in mass-loss rates and/or an increase in the SMBH feedback strength to preserve the current level of model-to-observation X-ray agreement.

\section*{Acknowledgements}

We thank the anonymous referee for helpful comments that significantly improved this paper.  We acknowledge M.\ A.\ Leutenegger for providing his code to tabulate X-ray opacities for given abundances and ionization states.  CMPR is supported by an appointment to the NASA Postdoctoral Program at the Goddard Space Flight Center, administered formerly by the Oak Ridge Associated Universities, and currently by the Universities Space Research Association, through a contract with NASA.  QDW acknowledges support by NASA via the SAO/CXC grant TM3-14006X.  JC acknowledges support from CONICYT-Chile through FONDECYT (1141175), Basal (PFB0609), and Anillo (ACT1101) grants.



\bibliographystyle{mnras}
\bibliography{sampleAF,sampleGM,sampleNS,sampleTZ}


\bsp	
\label{lastpage}
\end{document}